\documentclass[journal=jctc,manuscript=article]{achemso}
\setkeys{acs}{maxauthors=199, etalmode=truncate}
\usepackage[utf8]{inputenc}
\usepackage{textcomp}
\usepackage{graphicx}
\usepackage[version=4]{mhchem}
\usepackage{siunitx}
\DeclareSIUnit\au{a.u.}
\usepackage{tabularx}
\DeclareSIUnit\angstrom{\text{\AA}}

\usepackage{comment}
\usepackage{csquotes}
\usepackage{import}
\allowdisplaybreaks[1]
\MakeOuterQuote{"}
\usepackage{mathtools,tikz,caption}
\DeclareRobustCommand\sampleline[1]{%
  \tikz\draw[#1] (0,0) (0,\the\dimexpr\fontdimen22\textfont2\relax)
  -- (1.5em,\the\dimexpr\fontdimen22\textfont2\relax);%
}
\colorlet{review}{black}

\author{Tingting Zhao}
\affiliation{Department of Chemistry, Southern Methodist University, Dallas, TX}
\altaffiliation{These authors contributed equally to this work.}
\author{Megan Simons}
\affiliation{Department of Chemistry, Southern Methodist University, Dallas, TX}
\altaffiliation{These authors contributed equally to this work.}
\author{Devin A. Matthews}
\email{damatthews@smu.edu}
\affiliation{Department of Chemistry, Southern Methodist University, Dallas, TX}

\title{Open-shell Tensor Hypercontraction}

\abbreviations{THC, MP, LSTHC}

\begin{document}

\begin{tocentry}
\includegraphics{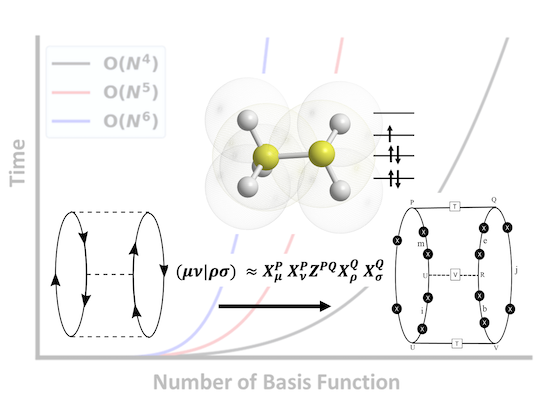}
\end{tocentry}

\begin{abstract}
The extension of least-squares tensor hypercontracted second- and third-order M{\o}ller-Plessett perturbation theory (LS-THC-MP2 and LS-THC-MP3) to open-shell systems is an important development due to the scaling reduction afforded by THC and the ubiquity of molecular ions, radicals, and other open-shell reactive species. The complexity of wavefunction-based quantum chemical methods such as M{\o}ller-Plessett and coupled cluster theory is reflected in the steep scaling of the computational costs with the molecular size. The least-squares tensor hypercontraction (LS-THC) method is an efficient, single-step factorization for the two-electron integral tensor, but can also be used to factorize the double excitation amplitudes, leading to significant scaling reduction. Here, we extend this promising method to open-shell variants of LS-THC-MP2 and -MP3 using diagrammatic techniques and explicit spin-summation. The accuracy of the resulting methods for open-shell species is benchmarked on standard tests systems such as regular alkanes, as well as realistic systems involving bond breaking, radical stabilization, and other effects. We find that open-shell LS-THC-MP$n$ methods exhibit errors highly comparable to those produced by closed-shell LS-THC-MP$n$, and are highly insensitive to particular chemical interactions, geometries, or even to moderate spin contamination.

\end{abstract}

\section{Introduction}

Complexity arising due to the presence of high-order tensors is well known as a bottleneck in wavefunction-based quantum chemical methods, leading to steep scaling of computational cost with system size. As the essential component of all electronic structure methods, the $4^{th}$-order electron repulsion integral (ERI) tensor, $g^{pq}_{rs} \equiv (pr|qs)$ (with the latter in chemists' notation) \cite{manybodymethods}, is the most obvious target for cost and scaling reduction. Many approaches have been developed to tackle the complexity of the ERIs, such as the density fitting (DF) approximation \cite{pewhitten, dunlap1979, Feyereisen1993, dunlapdf, vahtras1993} and similar techniques such as the resolution-of-the-identity (RI) \cite{riren, eichkorn1995, eichkorn1997, weigend2002, weigend2009}, Cholesky decomposition (CD) \cite{beebecholesky, kochcholesky, aquilantecholesky, weigend2009},  pseudospectral (PS) method \cite{martinezpseudospectral}, and more recently tensor hypercontraction (THC) \cite{thcmp2, thci, thcii, thciii, matthewslsthc2020, matthewslsthc2021}. Approximations of the wavefunction itself have received less attention, but due to the presence of the $4^{th}$-order double excitations amplitude ($\hat C_2$ or $\hat T_2$) which appear in virtually any wavefunction theory \cite{manybodymethods,ccbartlett}, some approximation is necessary to achieve significant reductions in computational scaling \cite{thciii, tensorcc, tensordecomp, tensordecompii,Parrish2019,hohenstein2022,schwerdtfeger,hoy2013,hoy2015}.

While a wide range of approximations to the ERIs and/or doubles amplitudes have proven highly successful, the vast majority of these methods have been implemented, tested on, and subsequently applied to closed-shell systems. While closed-shell systems are of vital importance to chemistry as a whole---neutral stable chemical species, both reactants and products, tend to adopt a closed-shell configuration---open-shell species are of equal importance due to the high reactivity and richness of electronic structure exhibited by radical species.\cite{krylovopenshell} Such species are central to a number of chemical fields, such as organic catalysis\cite{Radical_catalysis}, environmental and health studies\cite{Environmental_Xu2020}, combustion and alternative fuel chemistry\cite{Xing2019}, and photochemistry\cite{Zollitsch2018,SchweitzerChaput2018}. Additionally, due to the reactive nature of radical transient intermediates or transition states, they are often difficult to study in the laboratory, requiring more elaborate instrumentation\cite{Freitas2021} or complicated spectroscopic analysis \cite{Weckhuysen2015}. Thus, it is necessary to develop quantum chemistry methods to assist the experimental studies of these systems, along with approximation techniques to reduce computational scaling for the study of larger systems.

We are especially interested in the least squares tensor hypercontraction (LS-THC) \cite{thcii, matthewslsthc2020} method as an efficient approximation of both the ERI and doubles amplitudes tensors. We and others have previously demonstrated the potential for high accuracy and low scaling ($\mathcal{O}(N^4)$, where $N$ is a measure of system size) for closed shell third-order M{\o}ller-Plesset perturbation theory (MP3). \cite{leelinheadgordonthc, matthewslsthc2021} In LS-THC, the fourth order ERI tensor is approximated as a (hyper)product of five matrices: four collocation matrices, determined by evaluation of the molecular orbitals at a set of grid points, and a core matrix evaluated by least-squares fitting of the canonical ERIs (or ERIs approximated by a second method such as density fitting). In the "MP3a" variant, this approximation is combined with the Laplace transform quadrature technique \cite{Haser1992,constans2000,constans2003,kats2008} in order to compute the energy without referencing the doubles amplitudes explicitly. Alternatively, the "MP3b" method invokes an LS-THC factorization of the first-order doubles amplitudes and instead avoids the computational complexity of the Laplace transform quadrature (and a quadratic scaling with the number of quadrature points). Lastly, the "MP3c" and "MP3d" methods directly approximate the second-order doubles amplitudes, using an energy functional form identical to that used for coupled cluster and other more complex methods. \cite{ccbartlett,manybodymethods} This range of interpretations of MP3, as well as the basic building blocks of the MP3 energy (especially the well-known particle--particle ladder, hole--hole ladder, and "ring" terms), make MP3 an ideal stepping stone to more complete theories such as coupled cluster with single and double excitations (CCSD).

In this word, an extension of the LS-THC approach to open-shell MP2 and MP3 energies was developed using a mixed graphical-algebraic technique for deriving the working equations. We have implemented the MP3b variant of LS-THC-MP3, but the techniques developed are immediately applicable to the other MP3 variants as well as to more complex LS-THC methods.

\section{Theory}

The following notational conventions are used throughout this work:
\begin{itemize}
\item The letters $pqrs$ denote arbitrary molecular orbitals (MOs).
\item The letters $abcdef$ ($ijklmn$) denote virtual (occupied) MOs.
\item The letters $RSTUVWXY$ denote grid points.
\item The letters $JK$ denote auxiliary (density fitting) basis functions.
\item Where applicable, $\alpha$ ($\beta$) molecular orbitals, pruned grid points, and other quantities are denoted by an overbar (no overbar).
\end{itemize}

\subsection{Least-squares Tensor Hypercontraction}

Tensor hypercontraction (LS-THC) \cite{thci, thcii, thciii, matthewslsthc2020} is a method that combines the desirable features of several other factorizations: the representation of electron distributions over a linear-scaling auxiliary basis as in DF/RI/CD methods, the pseudospectral method's flexibility
of factoring exchange terms, the use of least-squares fitting as in DF and alternating least squares-based CP factorizations (although the LS-THC factorization is linear and non-iterative), and finally in the least-squares form a grid-based expansion as in the numerical integration of exchange--correlation functions in density functional theory. The LS-THC form of the ERIs is,
\begin{align}
(pr|qs) \equiv g_{rs}^{pq}\approx\sum_{RS}(X^{(pr)})_{p}^{R}(X^{(pr)})_{r}^{R}V_{RS}(X^{(qs)})_{q}^{S}(X^{(qs)})_{s}^{S} \label{eq:ls-thc}
\end{align}

The matrices $X$ are the collocation matrices, determined \emph{a priori} by evaluation of the (spatial) molecular orbitals $\psi_p$ at a set of grid points $r_R$: $X_{p}^{R}=\psi_{p}(r_{R})$. Superscripts such as $(pr)$ in \eqref{eq:ls-thc} differentiate different sets of "pruned" grid points.\cite{matthewslsthc2020} The pruning process is specific to the occupation of the pair of molecular orbitals, leading to separate occupied--occupied ($\mathbf{X} \equiv \mathbf{X}^{(ij)}$), mixed virtual--occupied ($\tilde{\mathbf{X}} \equiv \mathbf{X}^{(ai)} = \mathbf{X}^{(ia)}$), or virtual--virtual ($\tilde{\tilde{\mathbf{X}}} \equiv \mathbf{X}^{(ab)}$) collocation matrices. \textcolor{review}{The pruning procedure here is based on the Cholesky decomposition, but other pruning procedures are possible, e.g. based on domain decomposition.\cite{song_atomic_2016,song_atomic_2017}} These three cases may be further classified by the spin of the associated molecular orbitals (differentiated by an overbar on the MO and grid indices), leading to six unique collocation matrices for open-shell LS-THC-MP3.

The final matrix $\mathbf{V}$ is the core matrix, and is evaluated by least squares fitting of either the exact ERI tensor or some intermediate approximation. In this work we utilize density fitting in order to maintain an overall scaling of $\mathcal{O}(N^4)$\cite{matthewslsthc2021},
\begin{align}
\tilde{g}^{pr}_{qs}&=\sum_{JK}(pr|J)(J|K)^{-1}(K|qs) \\
V_{RS}&=\underset{V_{RS}}{\operatorname{argmin}}\,\frac{1}{2} \sum_{pqrs} \left|\tilde{g}^{pr}_{qs} - \sum_{RS} (X^{(pr)})_{p}^{R} (X^{(pr)})_{r}^{R} V_{RS} (X^{(qs)})_{q}^{S} (X^{(qs)})_{s}^{S} \right|^{2} \nonumber \\
&=\sum_{R'S'}(S^{(pr)}_{RR'})^{-1} E_{R'S'} (S^{(qs)}_{S'S})^{-1} \label{eq:least-squares} \\
E_{RS}&=\sum_{pqrs}(X^{(pr)})_{p}^{R}(X^{(pr)})_{r}^{R} \tilde{g}^{pr}_{qs} (X^{(qs)})_{q}^{S}(X^{(qs)})_{s}^{S} \label{eq:fitting} \\
S^{(pq)}_{RS}&=\sum_{pq}(X^{(pq)})_{p}^{R}(X^{(pq)})_{q}^{R}(X^{(pq)})_{p}^{S}(X^{(pq)})_{q}^{S}
\end{align}
where in each case a superscript $\bullet^{-1}$ is understood as an element of the matrix inverse rather than an inverse of the matrix element itself. Note that, as with the collocation matrices, there are as many as six distinct metric matrices $\mathbf{S}$. Due to the possible combination of each of these six electron distributions, there are as many as 21 different core ($\mathbf{V}$) and fitting ($\mathbf{E}$) matrices, of which 13 are utilized in open-shell LS-THC-MP3 (assuming a canonical Hartree--Fock reference for which the single excitation amplitudes can be neglected). In order to avoid notational clutter, we rely on context to determine which of these 13 core matrices \textcolor{review}{is} intended, unless explicitly specified.

\subsection{Canonical Formulation of MP2 and MP3\label{sec:mpn}}

In the canonical spin-orbital representation, the MP2 and MP3 energies are defined as,
\begin{align}
E_{MP2} = E_{MP2a} &= \frac{1}{2} \sum_{abij} \frac{(g^{ij}_{ab}-g^{ij}_{ba}) g^{ab}_{ij}}{\varepsilon_i+\varepsilon_j-\varepsilon_a-\varepsilon_b} \label{eq:mp2a} \\
E_{MP3} = E_{MP3a} &= \frac{1}{2} \sum_{abijkl} \frac{(g^{ij}_{ab}-g^{ij}_{ba}) g^{ij}_{kl} g^{ab}_{kl}}{(\varepsilon_i+\varepsilon_j-\varepsilon_a-\varepsilon_b)(\varepsilon_k+\varepsilon_l-\varepsilon_a-\varepsilon_b)} \nonumber \\
&+ \frac{1}{2} \sum_{abcdij} \frac{(g^{ij}_{ab}-g^{ij}_{ba}) g^{ab}_{cd} g^{cd}_{ij}}{(\varepsilon_i+\varepsilon_j-\varepsilon_a-\varepsilon_b)(\varepsilon_i+\varepsilon_j-\varepsilon_c-\varepsilon_d)}  \nonumber \\
&+ \sum_{abcijk} \frac{(g^{ij}_{ab}-g^{ij}_{ba}) (g^{bk}_{jc}-g^{bk}_{cj}) (g^{ac}_{ik}-g^{ac}_{ki})}{(\varepsilon_i+\varepsilon_j-\varepsilon_a-\varepsilon_b)(\varepsilon_i+\varepsilon_k-\varepsilon_a-\varepsilon_c)} \label{eq:mp3a} 
\end{align}
Note that the more common formulation of these energies uses the fully second-quantized representation of the Hamiltonian, $\hat{H} = \sum_{pq} f^p_q \{a_p^{\dagger} a_q\}_N + \frac{1}{4}\sum_{pqrs} v^{pq}_{rs} \{a_p^{\dagger} a_q^{\dagger} a_s a_r\}_N$, where $\{\bullet\}_N$ denotes normal-ordering. The two representations are connected by the simple identity $v^{pq}_{rs} = g^{pq}_{rs}-g^{pq}_{sr} = g^{pq}_{rs}-g^{qp}_{rs}$. The first-order and second-order perturbed double excitation amplitudes are most commonly defined by,
\begin{align}
t^{[n]}{}^{ab}_{ij} &= \check{t}^{[n]}{}^{ab}_{ij} - \check{t}^{[n]}{}^{ab}_{ji} = \check{t}^{[n]}{}^{ab}_{ij} - \check{t}^{[n]}{}^{ba}_{ij} \\
\check{t}^{[1]}{}^{ab}_{ij} &= \frac{g^{ab}_{ij}}{\varepsilon_i+\varepsilon_j-\varepsilon_a-\varepsilon_b} \label{eq:t2_1} \\
\check{t}^{[2]}{}^{ab}_{ij} &= \frac{\sum_{kl} g^{ij}_{kl} \check{t}^{[1]}{}^{ab}_{kl} + \sum_{cd} g^{ab}_{cd} \check{t}^{[1]}{}^{cd}_{ij} + \sum_{ck} (g^{bk}_{jc}-g^{bk}_{cj}) (\check{t}^{[1]}{}^{ac}_{ik}-\check{t}^{[1]}{}^{ac}_{ki})}{(\varepsilon_i+\varepsilon_j-\varepsilon_a-\varepsilon_b)} \label{eq:t2_2}
\end{align}
The representation of $\check{t}^{[n]}$ in a \emph{non-antisymmetric} form is critical in the following derivation, where we also show that such a representation is always well-defined by following a physically-motivated graphical derivation. With these definitions we can write several additional variant forms of the MP2 and MP3 energies, although all variants are mathematically identical when exact ERIs and perturbed amplitudes are used,
\begin{align}
E_{MP2b} &= \frac{1}{2} \sum_{abij} (g^{ij}_{ab}-g^{ij}_{ba}) \check{t}^{[1]}{}^{ab}_{ij} \\
E_{MP3b} &= \frac{1}{2} \sum_{abijkl} (\check{t}^{[1]\dagger}{}^{ij}_{ab}-\check{t}^{[1]\dagger}{}^{ij}_{ba}) g^{ij}_{kl} \check{t}^{[1]}{}^{ab}_{kl} + \frac{1}{2} \sum_{abcdij} (\check{t}^{[1]\dagger}{}^{ij}_{ab}-\check{t}^{[1]\dagger}{}^{ij}_{ba}) g^{ab}_{cd} \check{t}^{[1]}{}^{cd}_{ij}  \nonumber \\
&+ \sum_{abcijk} (\check{t}^{[1]\dagger}{}^{ij}_{ab}-\check{t}^{[1]\dagger}{}^{ij}_{ba}) (g^{bk}_{jc}-g^{bk}_{cj}) (\check{t}^{[1]}{}^{ac}_{ik}-\check{t}^{[1]}{}^{ac}_{ki}) \\
E_{MP3c} &= \frac{1}{2} \sum_{abij} (g^{ij}_{ab}-g^{ij}_{ba}) \check{t}^{[2]}{}^{ab}_{ij}
\end{align}
A further variant MP3d is obtained by separating the formation of $\check{t}^{[2]}$ into two parts: evaluation of the residual starting with the first-order amplitudes, and division of the residual by the orbital energy differences to obtain the second-order amplitudes. In Ref.~\citenum{matthewslsthc2021} we showed that this approach results in a distinct LS-THC-MP3 method with lower cost and nearly-identical numerical error.

Before moving on to the derivation of the LS-THC-MP$n$ approximations, we must deal with the inseparability of the energy denominators present in \eqref{eq:mp2a}--\eqref{eq:t2_2}. A convenient approach is the method of "Laplace denominators", pioneered by Almlöf and others, \cite{almlof1991,Haser1992}
\begin{align}
\frac{1}{\varepsilon_i+\varepsilon_j-\varepsilon_a-\varepsilon_b} &= -\int_{0}^{\infty} e^{(\varepsilon_{a}+\varepsilon_{b}-\varepsilon_{i}-\varepsilon_{j})t} dt \nonumber \\
&\approx -\sum_{\lambda=1}^{L}w_{\lambda}e^{(\varepsilon_{a}+\varepsilon_{b}-\varepsilon_{i}-\varepsilon_{j})t_{\lambda}} \nonumber \\
&= \sum_{\lambda=1}^{L}\tau_{a}^{\lambda}\tau_{b}^{\lambda}\tau_{i}^{\lambda}\tau_{j}^{\lambda} \label{eq:laplace}
\end{align}
where $L$ is the number of Laplace quadrature points. In this work we use the quadrature developed by Braess and Hackbusch\cite{braessApproximationExponentialSums2005} and a sufficient number of quadrature points to evaluate $1/x$ to a relative accuracy of $10^{-8}$ (approximately 9 points).

\subsection{Closed-Shell LS-THC-MP$n$}

The MP2a and MP3a formulations of MP2 and MP3, which are defined solely in terms of integrals and orbital energy denominators, are easily represented in compressed form using tensor hypercontraction. Spin-integration for closed-shell systems is straightforward (see e.g. Ref.~\citenum{manybodymethods}), and we may exploit the equivalence of the different spin-cases of the integrals,
\begin{align}
g^{pq}_{rs} = g^{\bar p\bar q}_{\bar r\bar s} = g^{p\bar q}_{r\bar s}
\end{align}
We may then expand the equations in terms of say, $g^{pq}_{rs}$ alone and then approximate these integrals using the form of \eqref{eq:ls-thc}, while also approximating the energy denominators using \eqref{eq:laplace}. This results in the LS-THC-MP2a and LS-THC-MP3a methods,
\begin{align}
E_{LS-THC-MP2a} &= - \sum_{abij} \sum_{RSTU} \sum_{\lambda=1}^L \tau_a^\lambda \tau_b^\lambda \tau_i^\lambda \tau_j^\lambda \left( 2 \tilde{X}_a^R \tilde{X}_i^R V_{RS} \tilde{X}_b^S \tilde{X}_j^S - \tilde{X}_a^R \tilde{X}_j^R V_{RS} \tilde{X}_b^S \tilde{X}_i^S \right) \nonumber \\
&\times \tilde{X}_a^T \tilde{X}_i^T V_{TU} \tilde{X}_b^U \tilde{X}_j^U \\
E_{LS-THC-MP3a} &= \sum_{abij} \sum_{RSTU} \sum_{\lambda=1}^L \tau_a^\lambda \tau_b^\lambda \tau_i^\lambda \tau_j^\lambda \left( 2 \tilde{X}_a^R \tilde{X}_i^R V_{RS} \tilde{X}_b^S \tilde{X}_j^S - \tilde{X}_a^R \tilde{X}_j^R V_{RS} \tilde{X}_b^S \tilde{X}_i^S \right) \nonumber \\
% ik,jl ak,bl
&\times \left[ \sum_{kl} \sum_{WY} \sum_{\lambda'=1}^L \tau_a^{\lambda'} \tau_b^{\lambda'} \tau_k^{\lambda'} \tau_l^{\lambda'} X_k^T X_i^T V_{TU} X_l^U X_j^U \tilde{X}_a^W \tilde{X}_k^W V_{WY} \tilde{X}_b^Y \tilde{X}_l^Y \right. \nonumber \\
% ac,bd ci,dj
& \left. + \sum_{cd} \sum_{WY} \sum_{\lambda'=1}^L \tau_c^{\lambda'} \tau_d^{\lambda'} \tau_i^{\lambda'} \tau_j^{\lambda'} \tilde{\tilde{X}}_a^T \tilde{\tilde{X}}_c^T V_{TU} \tilde{\tilde{X}}_b^U \tilde{\tilde{X}}_d^U \tilde{X}_c^W \tilde{X}_i^W V_{WY} \tilde{X}_d^Y \tilde{X}_j^Y \right. \nonumber \\
% 2 ai,kc ck,bj 
& \left. + 4 \sum_{ck} \sum_{WY} \sum_{\lambda'=1}^L \tau_a^{\lambda'} \tau_c^{\lambda'} \tau_i^{\lambda'} \tau_k^{\lambda'} \tilde{X}_c^T \tilde{X}_k^T V_{TU} \tilde{X}_b^U \tilde{X}_j^U \tilde{X}_a^W \tilde{X}_i^W V_{WY} \tilde{X}_c^Y \tilde{X}_k^Y \right. \nonumber \\
% - ai,kc cj,bk
& \left. - 2 \sum_{ck} \sum_{WY} \sum_{\lambda'=1}^L \tau_a^{\lambda'} \tau_c^{\lambda'} \tau_i^{\lambda'} \tau_k^{\lambda'} \tilde{X}_c^T \tilde{X}_k^T V_{TU} \tilde{X}_b^U \tilde{X}_j^U \tilde{X}_a^W \tilde{X}_k^W V_{WY} \tilde{X}_c^Y \tilde{X}_i^Y \right. \nonumber \\
% - ac,ki ck,bj
& \left. - 2 \sum_{ck} \sum_{WY} \sum_{\lambda'=1}^L \tau_a^{\lambda'} \tau_c^{\lambda'} \tau_i^{\lambda'} \tau_k^{\lambda'} \tilde{\tilde{X}}_b^T \tilde{\tilde{X}}_c^T V_{TU} X_j^U X_k^U \tilde{X}_a^W \tilde{X}_i^W V_{WY} \tilde{X}_c^Y \tilde{X}_k^Y \right. \nonumber \\
% - ac,kj ci,bk
& \left. - 2 \sum_{ck} \sum_{WY} \sum_{\lambda'=1}^L \tau_a^{\lambda'} \tau_c^{\lambda'} \tau_i^{\lambda'} \tau_k^{\lambda'} \tilde{\tilde{X}}_b^T \tilde{\tilde{X}}_c^T V_{TU} X_i^U X_k^U \tilde{X}_a^W \tilde{X}_k^W V_{WY} \tilde{X}_c^Y \tilde{X}_j^Y \right]
\end{align}
The "b" variants are simply obtained by directly approximating the first-order (orbital) doubles amplitudes using THC,
\begin{align}
t^{[1]}{}^{ab}_{ij} &= t^{[1]}{}^{\bar a\bar b}_{\bar i\bar j} = \check{t}^{[1]}{}^{ab}_{ij} - \check{t}^{[1]}{}^{ab}_{ji} \\
t^{[1]}{}^{a\bar b}_{i\bar j} &= \check{t}^{[1]}{}^{ab}_{ij} \approx \sum_{RS} \tilde{X}_a^R \tilde{X}_i^R T^{[1]}_{RS} \tilde{X}_b^S \tilde{X}_j^S \\
E_{LS-THC-MP2b} &= \sum_{abij} \sum_{RSTU} \left( 2 \tilde{X}_a^R \tilde{X}_i^R V_{RS} \tilde{X}_b^S \tilde{X}_j^S - \tilde{X}_a^R \tilde{X}_j^R V_{RS} \tilde{X}_b^S \tilde{X}_i^S \right) \nonumber \\
&\times \tilde{X}_a^T \tilde{X}_i^T T^{[1]}_{TU} \tilde{X}_b^U \tilde{X}_j^U \\
E_{LS-THC-MP3b} &= \sum_{abij} \sum_{RSTU} \left( 2 \tilde{X}_a^R \tilde{X}_i^R T^{[1]}_{RS} \tilde{X}_b^S \tilde{X}_j^S - \tilde{X}_a^R \tilde{X}_j^R T^{[1]}_{RS} \tilde{X}_b^S \tilde{X}_i^S \right) \nonumber \\
% ik,jl ak,bl
&\times \left[ \sum_{kl} \sum_{WY} X_k^T X_i^T V_{TU} X_l^U X_j^U \tilde{X}_a^W \tilde{X}_k^W T^{[1]}_{WY} \tilde{X}_b^Y \tilde{X}_l^Y \right. \nonumber \\
% ac,bd ci,dj
& \left. + \sum_{cd} \sum_{WY} \tilde{\tilde{X}}_a^T \tilde{\tilde{X}}_c^T V_{TU} \tilde{\tilde{X}}_b^U \tilde{\tilde{X}}_d^U \tilde{X}_c^W \tilde{X}_i^W T^{[1]}_{WY} \tilde{X}_d^Y \tilde{X}_j^Y \right. \nonumber \\
% 2 ai,kc ck,bj 
& \left. + 4 \sum_{ck} \sum_{WY} \tilde{X}_c^T \tilde{X}_k^T V_{TU} \tilde{X}_b^U \tilde{X}_j^U \tilde{X}_a^W \tilde{X}_i^W T^{[1]}_{WY} \tilde{X}_c^Y \tilde{X}_k^Y \right. \nonumber \\
% - ai,kc cj,bk
& \left. - 2 \sum_{ck} \sum_{WY} \tilde{X}_c^T \tilde{X}_k^T V_{TU} \tilde{X}_b^U \tilde{X}_j^U \tilde{X}_a^W \tilde{X}_k^W T^{[1]}_{WY} \tilde{X}_c^Y \tilde{X}_i^Y \right. \nonumber \\
% - ac,ki ck,bj
& \left. - 2 \sum_{ck} \sum_{WY} \tilde{\tilde{X}}_b^T \tilde{\tilde{X}}_c^T V_{TU} X_j^U X_k^U \tilde{X}_a^W \tilde{X}_i^W T^{[1]}_{WY} \tilde{X}_c^Y \tilde{X}_k^Y \right. \nonumber \\
% - ac,kj ci,bk
& \left. - 2 \sum_{ck} \sum_{WY} \tilde{\tilde{X}}_b^T \tilde{\tilde{X}}_c^T V_{TU} X_i^U X_k^U \tilde{X}_a^W \tilde{X}_k^W T^{[1]}_{WY} \tilde{X}_c^Y \tilde{X}_j^Y \right]
\end{align}
We do not further address the MP3c and MP3d variants and instead focus solely on LS-THC-MP3b.

These equations must be factorized (i.e. parentheses must be inserted to define to order of operations) before we can obtain efficient working equations. Here we adopt the same factorization as in our earlier work \cite{matthewslsthc2021}. As an example, consider the "particle-particle ladder" term of LS-THC-MP3b (depicted as the PP${}_C$ and PP${}_X$ diagrams in Fig.~\ref{fig:mp3_diagrams}),
\begin{align}
E_{PPLb} &= \sum_{abcdij} \sum_{RSTUWY} \left( 2 \tilde{X}_a^R \tilde{X}_i^R T^{[1]}_{RS} \tilde{X}_b^S \tilde{X}_j^S - \tilde{X}_a^R \tilde{X}_j^R T^{[1]}_{RS} \tilde{X}_b^S \tilde{X}_i^S \right) \nonumber \\
&\times \tilde{\tilde{X}}_a^T \tilde{\tilde{X}}_c^T V_{TU} \tilde{\tilde{X}}_b^U \tilde{\tilde{X}}_d^U \tilde{X}_c^W \tilde{X}_i^W T^{[1]}_{WY} \tilde{X}_d^Y \tilde{X}_j^Y \nonumber \\
&= 2 \sum_{Sbj} \left( \sum_U \left( \sum_Y \left( \sum_c \left( \sum_a \tilde{\tilde{P}}_{ac}^U \tilde{X}_a^S \right) \left( \sum_i Q_{ci}^Y \tilde{X}_i^S \right) \right) \left( \sum_d \tilde{\tilde{X}}_d^U \tilde{X}_d^Y \right) \tilde{X}_j^Y \right) \tilde{\tilde{X}}_b^U \right) Q_{bj}^S \nonumber \\
&- \sum_{STY} \left( \sum_d \left( \sum_b \tilde{\tilde{P}}_{bd}^T \tilde{X}_b^S \right) \tilde{X}_d^Y \right) \left( \sum_i \left( \sum_c Q_{ci}^Y \tilde{\tilde{X}}_c^T \right) \tilde{X}_i^S \right) \left( \sum_j \left( \sum_a Q_{aj}^S \tilde{\tilde{X}}_a^T \right) \tilde{X}_j^Y \right) \\
\tilde{\tilde{P}}^R_{ab} &= \sum_S V_{RS} \tilde{\tilde{X}}_a^S \tilde{\tilde{X}}_b^S \\
Q_{ai}^R &= \sum_S T^{[1]}_{RS} \tilde{X}_a^S \tilde{X}_i^S
\end{align}
This factorization enables each step to be completed in at most $\mathcal{O}(N^4)$ time, given a linear number of occupied and virtual orbitals as well as grid points. The full working equations for the closed-shell case are given in Ref.~\citenum{matthewslsthc2021}, and those for the open-shell case are given in the ESI.

\subsection{Open-Shell LS-THC-MP$n$}

\begin{figure}[!h]
    \centering
    \includegraphics[width=0.4\textwidth]{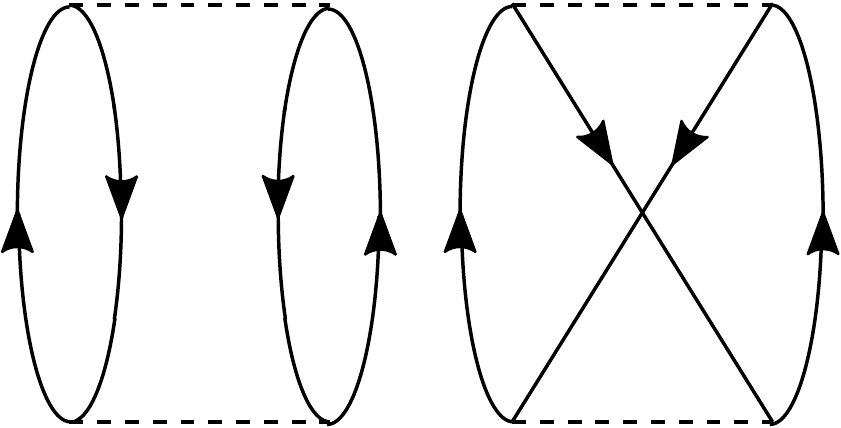}
    \caption{Goldstone diagrams for the MP2 energy, omitting denominator lines for clarity.}
    \label{fig:mp2_diagrams}
\end{figure}

\begin{figure}[!h]
    \centering
    \includegraphics[width=0.8\textwidth]{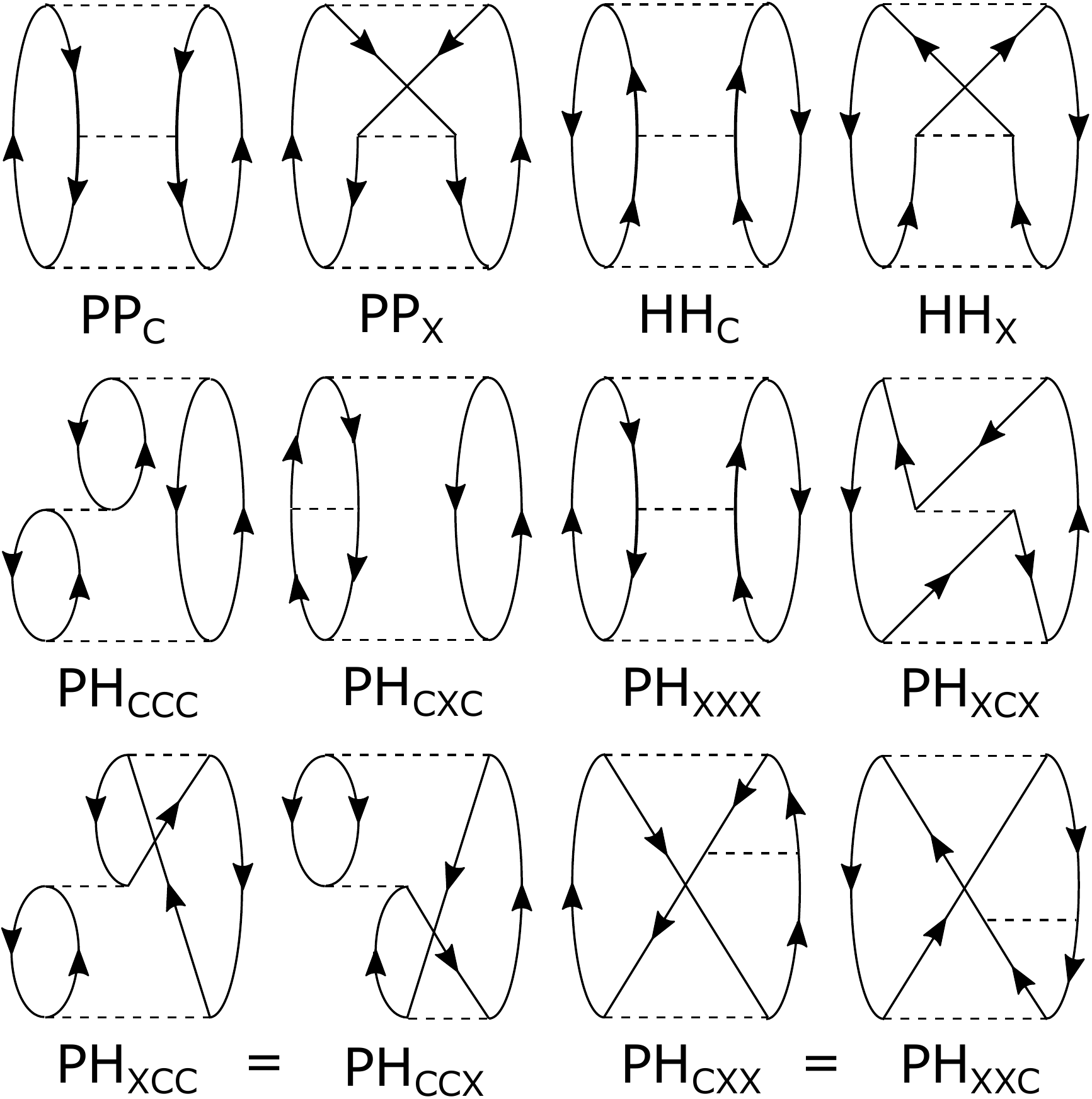}
    \caption{Goldstone diagrams for the MP3 energy, omitting denominator lines for clarity.}
    \label{fig:mp3_diagrams}
\end{figure}

\begin{figure}[!h]
    \centering
    \includegraphics[width=0.8\textwidth]{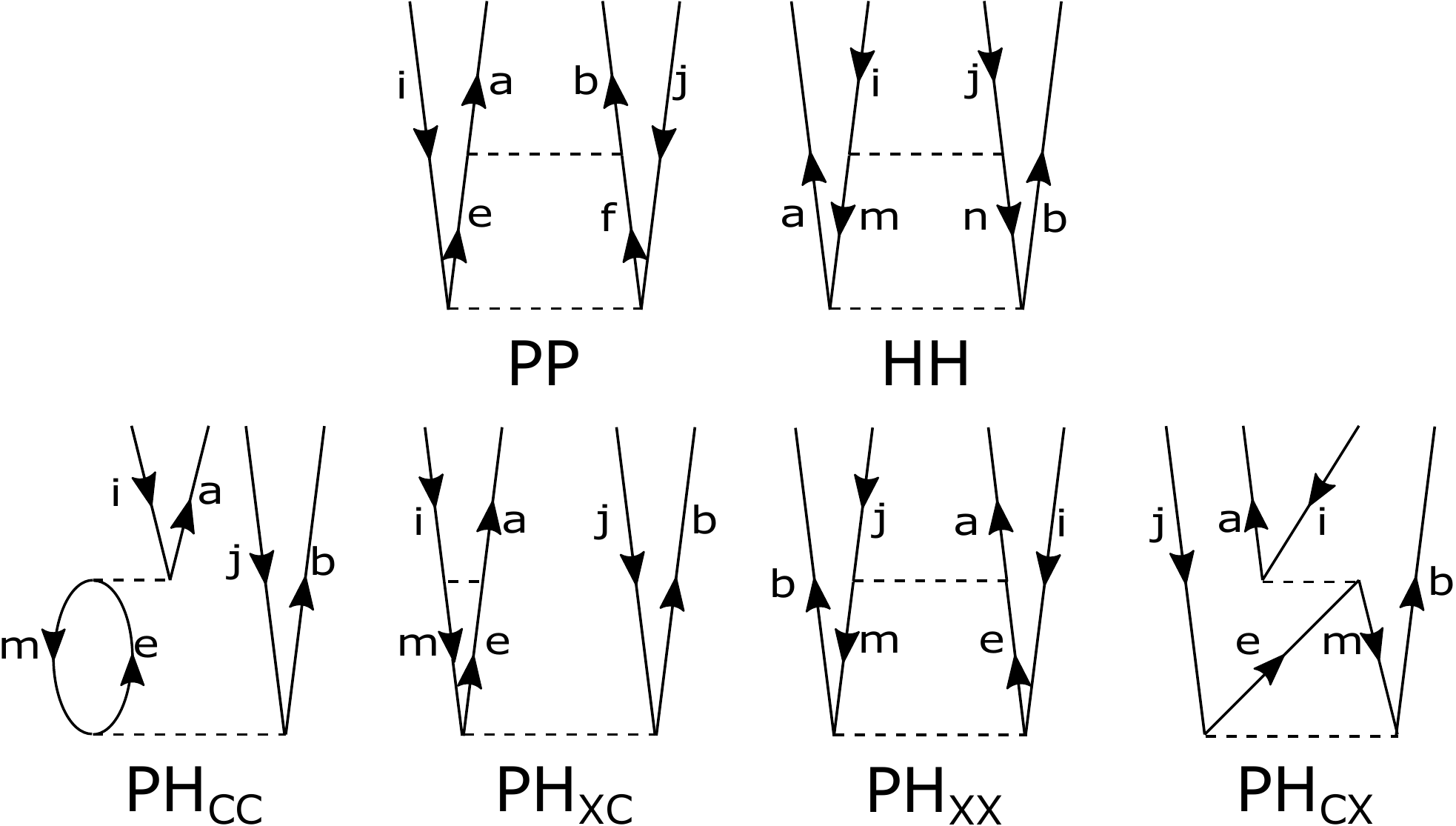}
    \caption{Goldstone diagrams for the $\check{t}^{[2]}{}^{ab}_{ij}$ amplitudes, omitting denominator lines for clarity.}
    \label{fig:t2_diagrams}
\end{figure}

In the preceding section, we used the non-antisymmetrized doubles amplitudes $\check{t}^{[n]}$. We can justify the validity of such a representation in the open-shell case and utilize it to easily derive open-shell variants of LS-THC-MP$n$ methods by utilizing diagrammatic methods.

First, we may recognize each term in \eqref{eq:mp2a}, \eqref{eq:mp3a}, and \eqref{eq:t2_2}, after distribution of the parentheses, as a unique Goldstone diagram.\cite{matthewslsthc2021} \textcolor{review}{Here, we use (non-antisymmetrized) Goldstone diagrams instead of the usual antisymetrized Goldstone diagrams (ASGs), also called Brandow diagrams.\cite{manybodymethods,ccbartlett} As will be seen below, avoiding explicit antisymmetrization allows the same-spin amplitudes and integrals to be factorized in way which avoids factorization of "exchange-like" terms. The unique Goldstone} diagrams are reproduced in Figures~\ref{fig:mp2_diagrams}, \ref{fig:mp3_diagrams}, and \ref{fig:t2_diagrams}, respectively. In particular, the diagrams in Figure~\ref{fig:t2_diagrams} are easily recognized as the necessary contributions to the closed-shell second-order $\hat{T}$ amplitudes or, with replacement of the bottom integral vertex by doubles amplitudes, the iterative $\hat{T}_2 \rightarrow \hat{T}_2$ contributions in coupled cluster theory. In the closed-shell interpretation, closed loops contribute a factor of 2 and each diagram is symmetrized, leading to the well-known expression,
\begin{align}
(\varepsilon_a + \varepsilon_b - \varepsilon_i - \varepsilon_j) \check{t}^{[2]}{}^{ab}_{ij}
&= \sum_{ef} g^{ab}_{ef} \check{t}^{[1]}{}^{ef}_{ij}
+ \sum_{mn} g^{mn}_{ij} \check{t}^{[1]}{}^{ab}_{mn}
+ (1+P^{ai}_{bj}) \left( \sum_{em} 2 g^{am}_{ie} \check{t}^{[1]}{}^{eb}_{mj} \right. \nonumber \\
&- \left. \sum_{em} g^{am}_{ie} \check{t}^{[1]}{}^{eb}_{jm}
- \sum_{em} g^{am}_{ei} \check{t}^{[1]}{}^{eb}_{mj}
- \sum_{em} g^{am}_{ej} \check{t}^{[1]}{}^{eb}_{im} \right)
\end{align}
where the permutation operator $P^{ai}_{bj}$ exchanges the top labels with those on the bottom in the following expression. Note that in the last term on the right-hand side, the $ij$ orbitals are ordered differently than in the remaining terms. This highlights the rule for Goldstone diagrams that, for orbitals sharing the same "column" in the external vertex (e.g. $ai$ or $bj$), we must be able to follow a continuous loop through the diagram from one label to the other. Thus the "PH${}_{XX}$" diagram of Fig.~\ref{fig:t2_diagrams} results in such a modified labeling. This choice is discussed further and theoretically motivated below.

For the closed-shell case we may use $\check{t}^{ab}_{ij}$ without ambiguity since it is precisely defined as the mixed-spin amplitudes $t^{a\bar{b}}_{i\bar{j}}$ due to the relationship between the various spin cases in the closed-shell case.  \cite{manybodymethods} However, in the open-shell case we could instead replace the factor of 2 for closed loops by an explicit summation over spin. In any case, each contraction line along a loop must carry the same spin since contraction implies a spin integral over orthonormal spin functions. Then, we can derive equations for the three distinct spin cases in the spin-unrestricted formalism,
\begin{align}
(\varepsilon_a + \varepsilon_b - \varepsilon_i - \varepsilon_j) \check{t}^{[2]}{}^{ab}_{ij}
&= \sum_{ef} g^{ab}_{ef} \check{t}^{[1]}{}^{ef}_{ij}
+ \sum_{mn} g^{mn}_{ij} \check{t}^{[1]}{}^{ab}_{mn}
+ (1+P^{ai}_{bj}) \left( \sum_{em} g^{am}_{ie} \check{t}^{[1]}{}^{eb}_{mj} 
 \right. \nonumber \\
&+ \sum_{e\bar m} g^{a\bar m}_{i\bar e} \check{t}^{[1]}{}^{\bar eb}_{\bar mj}
- \left. \sum_{em} g^{am}_{ie} \check{t}^{[1]}{}^{eb}_{jm}
- \sum_{em} g^{am}_{ei} \check{t}^{[1]}{}^{eb}_{mj}
- \sum_{em} g^{am}_{ej} \check{t}^{[1]}{}^{eb}_{im} \right) \label{eq:t2_2_aa}\\
(\varepsilon_a + \varepsilon_{\bar b} - \varepsilon_i - \varepsilon_{\bar j}) \check{t}^{[2]}{}^{a\bar b}_{i\bar j}
&= \sum_{e\bar f} g^{a\bar b}_{e\bar f} \check{t}^{[1]}{}^{e\bar f}_{i\bar j}
+ \sum_{m\bar n} g^{m\bar n}_{i\bar j} \check{t}^{[1]}{}^{a\bar b}_{m\bar n}
+ \sum_{em} g^{am}_{ie} \check{t}^{[1]}{}^{e\bar b}_{m\bar j}
+ \sum_{\bar e\bar m} g^{a\bar m}_{i\bar e} \check{t}^{[1]}{}^{\bar e\bar b}_{\bar m\bar j} \nonumber \\
&+ \sum_{em} g^{\bar bm}_{\bar je} \check{t}^{[1]}{}^{ea}_{mi}
+ \sum_{\bar e\bar m} g^{\bar b\bar m}_{\bar j\bar e} \check{t}^{[1]}{}^{\bar ea}_{\bar mi}
- \sum_{\bar e\bar m} g^{a\bar m}_{i\bar e} \check{t}^{[1]}{}^{\bar e\bar b}_{\bar j\bar m}
- \sum_{em} g^{am}_{ei} \check{t}^{[1]}{}^{e\bar b}_{m\bar j} \nonumber \\
&- \sum_{e\bar m} g^{a\bar m}_{e\bar j} \check{t}^{[1]}{}^{e\bar b}_{i\bar m}
- \sum_{em} g^{\bar bm}_{\bar je} \check{t}^{[1]}{}^{ea}_{im}
- \sum_{\bar e\bar m} g^{\bar b\bar m}_{\bar e\bar j} \check{t}^{[1]}{}^{\bar ea}_{\bar mi}
- \sum_{\bar em} g^{\bar bm}_{\bar ei} \check{t}^{[1]}{}^{\bar ea}_{\bar jm} \label{eq:t2_2_ab}\\
(\varepsilon_{\bar a} + \varepsilon_{\bar b} - \varepsilon_{\bar i} - \varepsilon_{\bar j}) \check{t}^{[2]}{}^{\bar a\bar b}_{\bar i\bar j}
&= \sum_{\bar e\bar f} g^{\bar a\bar b}_{\bar e\bar f} \check{t}^{[1]}{}^{\bar e\bar f}_{\bar i\bar j}
+ \sum_{\bar m\bar n} g^{\bar m\bar n}_{\bar i\bar j} \check{t}^{[1]}{}^{\bar a\bar b}_{\bar m\bar n}
+ (1+P^{\bar a\bar i}_{\bar b\bar j}) \left( \sum_{\bar e\bar m} g^{\bar a\bar m}_{\bar i\bar e} \check{t}^{[1]}{}^{\bar e\bar b}_{\bar m\bar j} 
 \right. \nonumber \\
&+ \sum_{\bar em} g^{\bar am}_{\bar ie} \check{t}^{[1]}{}^{e\bar b}_{m\bar j}
- \left. \sum_{\bar e\bar m} g^{\bar a\bar m}_{\bar i\bar e} \check{t}^{[1]}{}^{\bar e\bar b}_{\bar j\bar m}
- \sum_{\bar e\bar m} g^{\bar a\bar m}_{\bar e\bar i} \check{t}^{[1]}{}^{\bar e\bar b}_{\bar m\bar j}
- \sum_{\bar e\bar m} g^{\bar a\bar m}_{\bar e\bar j} \check{t}^{[1]}{}^{\bar e\bar b}_{\bar i\bar m} \right) \label{eq:t2_2_bb}
\end{align}
Antisymmetrization of \eqref{eq:t2_2_aa} and \eqref{eq:t2_2_bb} in accordance with our definition of $\check{t}^{[2]}{}^{ab}_{ij}$ arrives precisely at the standard equations for the second-order amplitudes in a (canonical) unrestricted Hartree--Fock reference. The application of Goldstone diagrams along with explicit spin-summation then gives us a rather simple route to derive the open-shell working equations. \textcolor{review}{The use of Goldstone diagrams also offers a straightforward way to implement THC for spin-component-scaled Møller–Plesset (SCS-MP) perturbation theories\cite{scsmp2,song_atomic_2016,song_atomic_2017, song_analytical_2017}, by simply including a scaling coefficient in each spin-labeled Goldstone diagram.}
The necessary equivalence of these equations with the standard spin-integrated form \emph{after} antisymmetrization, and that these equations trivially reduce to the one given above in the closed-shell is what we mean by claiming that $\check{t}^{[2]}$ is well-defined. As we discuss below, however, these amplitudes and their factorized form are not numerically well-defined, but we can make a consistent choice based on theoretical arguments.

A recursive application of \eqref{eq:t2_2_aa}--\eqref{eq:t2_2_bb}, with additional terms accounting for single excitation amplitudes, provides a route to define non-antisymmetrized amplitudes for methods such as coupled cluster with single and double excitations (CCSD). In the present work, we focus on LS-THC-MP3b, where in fact we only require $\check{t}^{[1]}$. These amplitudes are trivially defined in terms of the orbital two-electron integrals,
\begin{align}
\check{t}^{[1]}{}^{ab}_{ij} = \frac{g^{ab}_{ij}}{\epsilon_a + \epsilon_b - \epsilon_i - \epsilon_j} \approx \sum_{RS} \tilde{X}_a^R \tilde{X}_i^R T^{[1]}_{RS} \tilde{X}_b^S \tilde{X}_j^S \\
\check{t}^{[1]}{}^{a\bar b}_{i\bar j} = \frac{g^{a\bar b}_{i\bar j}}{\epsilon_a + \epsilon_{\bar b} - \epsilon_i - \epsilon_{\bar j}} \approx \sum_{R\bar S} \tilde{X}_a^R \tilde{X}_i^R T^{[1]}_{R\bar S} \tilde{X}_{\bar b}^{\bar S} \tilde{X}_{\bar j}^{\bar S} \\
\check{t}^{[1]}{}^{\bar a\bar b}_{\bar i\bar j} = \frac{g^{\bar a\bar b}_{\bar i\bar j}}{\epsilon_{\bar a} + \epsilon_{\bar b} - \epsilon_{\bar i} - \epsilon_{\bar j}} \approx \sum_{\bar R\bar S} \tilde{X}_{\bar a}^{\bar R} \tilde{X}_{\bar i}^{\bar R} T^{[1]}_{\bar R\bar S} \tilde{X}_{\bar b}^{\bar S} \tilde{X}_{\bar j}^{\bar S}
\end{align}
We note that these chosen definitions, while perhaps obvious given the close relationship of the first-order amplitudes to the integrals, are also derivable using the diagrammatic technique outlined above.

\definecolor{lightgreen}{RGB}{141,211,95}
\definecolor{darkgreen}{RGB}{0,128,0}
\definecolor{darkorange}{RGB}{255,127,42}
\definecolor{lightorange}{RGB}{255,212,42}
\definecolor{pinkish}{RGB}{255,42,127}
\begin{figure}[!h]
    \centering
    \includegraphics[width=0.8\textwidth]{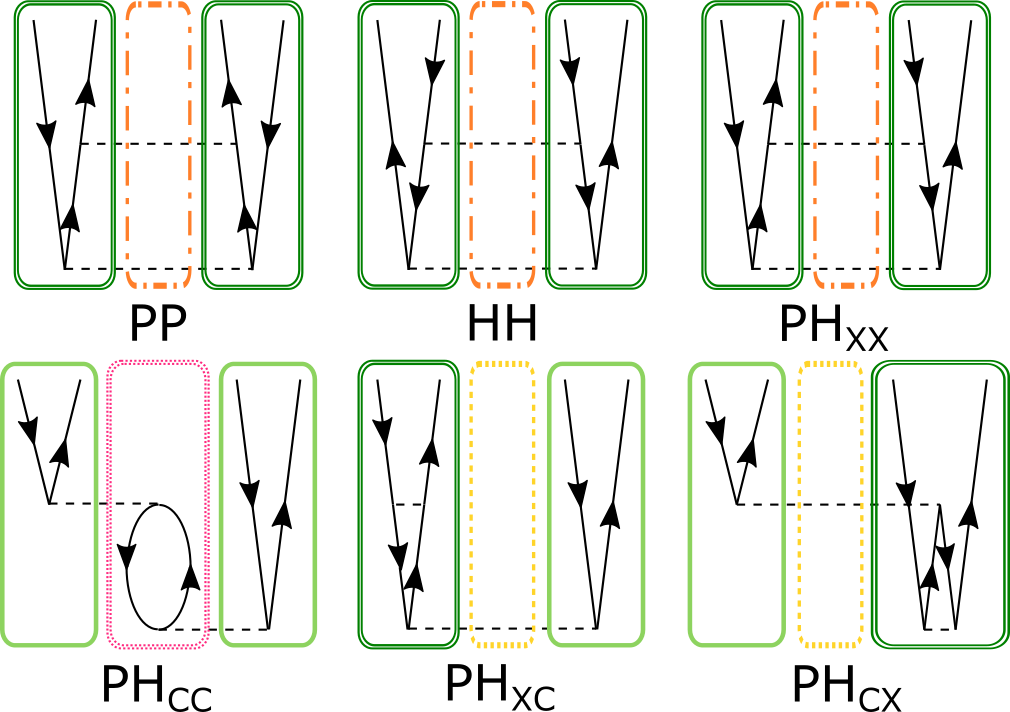}
    \caption{Goldstone diagrams for the $\check{t}^{[2]}{}^{ab}_{ij}$ amplitudes, highlighting the effect of the THC factorization. Particle-hole pairs, in solid lines, are represented via projection form the orbital to the grid space using pairs of collocation matrices $\mathbf{X}$. Interactions, in dashed/dotted lines are represented by the core matrix $\mathbf{V}^{[2]}$. The second-order structure leads to multiple types of particle-hole pairs and inter-pair interactions: \sampleline{very thick, color=lightgreen}particle-hole pair, \sampleline{thick, double distance=1pt, color=darkgreen}dressed particle-hole pair, \sampleline{very thick, dotted, color=lightorange}bare interaction, \sampleline{very thick, dash dot, color=darkorange}bare interaction in Cartesian product space, \sampleline{thick, double distance=1pt, dotted, color=pinkish}dressed interaction.}
    \label{fig:t2_diagrams_factorized}
\end{figure}

Returning to the issue of relabeling the $ij$ indices in the ring terms, we do note that for the same-spin amplitudes, \emph{every} term may in fact be written in one of two ways due to the $P(ij)$ [or equivalently $P(ab)$] factor which relates the antisymmetric and non-antisymmetric amplitudes. Thus, we may technically define a large number of equivalent non-antisymmetric amplitudes. However, once a tensor factorization such as THC is applied, these different definitions are no longer equivalent and may incur errors of differing magnitude. In this regard, our specific choice of \eqref{eq:t2_2_aa} and \eqref{eq:t2_2_bb} is motivated by the physical compressibility (in an information-theoretic sense) of inter-electron interactions. The "same column same loop" rule espoused above allows us to interpret a tensor factorization of the entire diagram as a "recompression" step. In Fig.~\ref{fig:t2_diagrams_factorized} we conceptually identify regions of each Goldstone diagram with the resulting factors in the THC compressed form: electron pairs will be represented by a pair of collocation matrices e.g. $\tilde{X}_a^R \tilde{X}_i^R$ which effect a transformation from molecular orbital to grid space, and pair correlation is represented by the core matrix $\mathbf{T}^{[2]}$. The orbital labeling in Fig.~\ref{fig:t2_diagrams_factorized} demands that the collocation matrices capture some correlation effects leading to "dressed" particle-hole pairs (which contributes to the numerical errors discussed previously by us\cite{matthewslsthc2021}), while the interaction must capture bare, dressed, and higher-dimensional interactions via a linear-scaling extraction of important features, in a similar spirit to the singular value decomposition and other low-rank decompositions. Switching either the $ij$ or $ab$ labels destroys this clear separation in terms of local (raw or dressed Coulombic) interactions of dressed particle-hole pairs, and instead introduces a picture with more exchange-like interactions. The THC decomposition, like most low-rank decompositions, is not able to effectively capture such non-local interactions.

A final numerical issue for the $\check{t}^{ab}_{ij}$ and $\check{t}^{\bar a\bar b}_{\bar i\bar j}$ amplitudes are the exclusion principle-violating (EPV) amplitudes which occur when $i=j$ or $a=b$. In the canonical equations these amplitudes cancel after antisymmetrization and do not affect the energy or properties in any way. Thus, we could assign these amplitudes any numerical value. However, when applying the tensor hypercontraction approximation, information from \emph{all} doubles amplitudes is mixed together to determine the elements of the core matrix $\mathbf{T}^{[2]}$ via least-squares fitting. Thus, the EPV amplitudes do potentially contribute to the LS-THC-MP3 energy. We argue that, in order to minimize the impact of EPV amplitudes, we should choose a value for these amplitudes which produces the most accurate THC decomposition of the doubles amplitudes as a whole. This guideline is based on the fact that as the THC decomposition approaches exactness (e.g. as the grid size is increased), the cancellation of the EPV terms, regardless of their numerical value, becomes more complete. This approach is quite different from setting the EPV amplitudes themselves to the smallest possible value (namely, zero). Thus, we suggest to not modify the EPV terms from their definition as obtained using the above diagrammatic approach, e.g. \eqref{eq:t2_2_aa}--\eqref{eq:t2_2_bb}. Because the THC approximation captures the global mathematical structure of the amplitudes (i.e. it is an interpolation), a consistent choice of EPV and non-EPV terms should provide the most compressible amplitudes. For $\check{t}^{[1]}$, which are the only amplitudes used in LS-THC-MP3b, the proposed choice of the EPV amplitudes also corresponds directly to the "correct" Coulomb self-interaction.

Following the diagrammatic method using Fig.~\ref{fig:mp3_diagrams} (where we replace the top and bottom Hamiltonian vertices by the first-order amplitudes), followed by THC approximation of the integrals and first-order amplitudes for each spin case, we arrive at equations for open-shell LS-THC-MP3b,
\begin{align}
E_{LS-THC-MP3b} &= \sum_{abij} \sum_{RS} \left( \tilde{X}_a^R \tilde{X}_i^R T^{[1]}_{RS} \tilde{X}_b^S \tilde{X}_j^S - \tilde{X}_a^R \tilde{X}_b^R T^{[1]}_{RS} \tilde{X}_b^S \tilde{X}_i^S \right) \nonumber \\
% IK,JL AK,BL
&\times \left[ \frac{1}{2} \sum_{kl} \sum_{TUWY} X_k^T X_i^T V_{TU} X_l^U X_j^U \tilde{X}_a^W \tilde{X}_k^W T^{[1]}_{WY} \tilde{X}_b^Y \tilde{X}_l^Y \right. \nonumber \\
% AC,BD CI,DJ
& \left. + \frac{1}{2} \sum_{cd} \sum_{TUWY} \tilde{\tilde{X}}_a^T \tilde{\tilde{X}}_c^T V_{TU} \tilde{\tilde{X}}_b^U \tilde{\tilde{X}}_d^U \tilde{X}_c^W \tilde{X}_i^W T^{[1]}_{WY} \tilde{X}_d^Y \tilde{X}_j^Y \right. \nonumber \\
% AI,KC CK,BJ
& \left. + \sum_{ck} \sum_{TUWY} \tilde{X}_c^T \tilde{X}_k^T V_{TU} \tilde{X}_b^U \tilde{X}_j^U \tilde{X}_a^W \tilde{X}_i^W T^{[1]}_{WY} \tilde{X}_c^Y \tilde{X}_k^Y \right. \nonumber \\
% AI,kc ck,BJ
& \left. + \sum_{\bar c\bar k} \sum_{\bar TUW\bar Y} \tilde{X}_{\bar c}^{\bar T} \tilde{X}_{\bar k}^{\bar T} V_{\bar TU} \tilde{X}_b^U \tilde{X}_j^U \tilde{X}_a^W \tilde{X}_i^W T^{[1]}_{W\bar Y} \tilde{X}_{\bar c}^{\bar Y} \tilde{X}_{\bar k}^{\bar Y} \right. \nonumber \\
% - AI,KC CJ,BK
& \left. - \sum_{ck} \sum_{TUWY} \tilde{X}_c^T \tilde{X}_k^T V_{TU} \tilde{X}_b^U \tilde{X}_j^U \tilde{X}_a^W \tilde{X}_k^W T^{[1]}_{WY} \tilde{X}_c^Y \tilde{X}_i^Y \right. \nonumber \\
% - AC,KI CK,BJ
& \left. - \sum_{ck} \sum_{TUWY} \tilde{\tilde{X}}_b^T \tilde{\tilde{X}}_c^T V_{TU} X_j^U X_k^U \tilde{X}_a^W \tilde{X}_i^W T^{[1]}_{WY} \tilde{X}_c^Y \tilde{X}_k^Y \right. \nonumber \\
% - AC,KJ CI,BK
& \left. - \sum_{ck} \sum_{TUWY} \tilde{\tilde{X}}_b^T \tilde{\tilde{X}}_c^T V_{TU} X_i^U X_k^U \tilde{X}_a^W \tilde{X}_k^W T^{[1]}_{WY} \tilde{X}_c^Y \tilde{X}_j^Y \right] \nonumber \\
&+ \sum_{a\bar bi\bar j} \sum_{R\bar S} \tilde{X}_a^R \tilde{X}_i^R T^{[1]}_{R\bar S} \tilde{X}_{\bar b}^{\bar S} \tilde{X}_{\bar j}^{\bar S} \nonumber \\
% IK,JL AK,BL
&\times \left[ \sum_{k\bar l} \sum_{T\bar UW\bar Y} X_k^T X_i^T V_{T\bar U} X_{\bar l}^{\bar U} X_{\bar j}^{\bar U} \tilde{X}_a^W \tilde{X}_k^W T^{[1]}_{W\bar Y} \tilde{X}_{\bar b}^{\bar Y} \tilde{X}_{\bar l}^{\bar Y} \right. \nonumber \\
% AC,BD CI,DJ
& \left. + \sum_{c\bar d} \sum_{T\bar UW\bar Y} \tilde{\tilde{X}}_a^T \tilde{\tilde{X}}_c^T V_{T\bar U} \tilde{\tilde{X}}_{\bar b}^{\bar U} \tilde{\tilde{X}}_{\bar d}^{\bar U} \tilde{X}_c^W \tilde{X}_i^W T^{[1]}_{W\bar Y} \tilde{X}_{\bar d}^{\bar Y} \tilde{X}_{\bar j}^{\bar Y} \right. \nonumber \\
% AI,KC CK,BJ
& \left. + \sum_{ck} \sum_{T\bar UWY} \tilde{X}_c^T \tilde{X}_k^T V_{T\bar U} \tilde{X}_{\bar b}^{\bar U} \tilde{X}_{\bar j}^{\bar U} \tilde{X}_a^W \tilde{X}_i^W T^{[1]}_{WY} \tilde{X}_c^Y \tilde{X}_k^Y \right. \nonumber \\
% AI,kc ck,BJ
& \left. + \sum_{\bar c\bar k} \sum_{\bar T\bar UW\bar Y} \tilde{X}_{\bar c}^{\bar T} \tilde{X}_{\bar k}^{\bar T} V_{\bar T\bar U} \tilde{X}_{\bar b}^{\bar U} \tilde{X}_{\bar j}^{\bar U} \tilde{X}_a^W \tilde{X}_i^W T^{[1]}_{W\bar Y} \tilde{X}_{\bar c}^{\bar Y} \tilde{X}_{\bar k}^{\bar Y} \right. \nonumber \\
% - AI,KC CJ,BK
& \left. - \sum_{ck} \sum_{T\bar UWY} \tilde{X}_c^T \tilde{X}_k^T V_{T\bar U} \tilde{X}_{\bar b}^{\bar U} \tilde{X}_{\bar j}^{\bar U} \tilde{X}_a^W \tilde{X}_k^W T^{[1]}_{WY} \tilde{X}_c^Y \tilde{X}_i^Y \right. \nonumber \\
% - AC,KI CK,BJ
& \left. - \sum_{\bar c\bar k} \sum_{\bar T\bar UW\bar Y} \tilde{\tilde{X}}_{\bar b}^{\bar T} \tilde{\tilde{X}}_{\bar c}^{\bar T} V_{\bar T\bar U} X_{\bar j}^{\bar U} X_{\bar k}^{\bar U} \tilde{X}_a^W \tilde{X}_i^W T^{[1]}_{W\bar Y} \tilde{X}_{\bar c}^{\bar Y} \tilde{X}_{\bar k}^{\bar Y} \right. \nonumber \\
% - AC,KJ CI,BK
& \left. - \sum_{\bar ck} \sum_{\bar TUW\bar Y} \tilde{\tilde{X}}_{\bar b}^{\bar T} \tilde{\tilde{X}}_{\bar c}^{\bar T} V_{\bar TU} X_i^U X_k^U \tilde{X}_a^W \tilde{X}_k^W T^{[1]}_{W\bar Y} \tilde{X}_{\bar c}^{\bar Y} \tilde{X}_{\bar j}^{\bar Y} \right. \nonumber \\
% BJ,KC CK,AI
& \left. + \sum_{ck} \sum_{TU\bar WY} \tilde{X}_c^T \tilde{X}_k^T V_{TU} \tilde{X}_a^U \tilde{X}_i^U \tilde{X}_{\bar b}^{\bar W} \tilde{X}_{\bar j}^{\bar W} T^{[1]}_{\bar WY} \tilde{X}_c^Y \tilde{X}_k^Y \right. \nonumber \\
% BJ,kc ck,AI
& \left. + \sum_{\bar c\bar k} \sum_{\bar TU\bar W\bar Y} \tilde{X}_{\bar c}^{\bar T} \tilde{X}_{\bar k}^{\bar T} V_{\bar TU} \tilde{X}_a^U \tilde{X}_i^U \tilde{X}_{\bar b}^{\bar W} \tilde{X}_{\bar j}^{\bar W} T^{[1]}_{\bar W\bar Y} \tilde{X}_{\bar c}^{\bar Y} \tilde{X}_{\bar k}^{\bar Y} \right. \nonumber \\
% - BJ,KC CI,AK
& \left. - \sum_{\bar c\bar k} \sum_{\bar TU\bar W\bar Y} \tilde{X}_{\bar c}^{\bar T} \tilde{X}_{\bar k}^{\bar T} V_{\bar TU} \tilde{X}_a^U \tilde{X}_i^U \tilde{X}_{\bar b}^{\bar W} \tilde{X}_{\bar k}^{\bar W} T^{[1]}_{\bar W\bar Y} \tilde{X}_{\bar c}^{\bar Y} \tilde{X}_{\bar j}^{\bar Y} \right. \nonumber \\
% - BC,KJ CK,AI
& \left. - \sum_{ck} \sum_{TU\bar WY} \tilde{\tilde{X}}_a^T \tilde{\tilde{X}}_c^T V_{TU} X_i^U X_k^U \tilde{X}_{\bar b}^{\bar W} \tilde{X}_{\bar j}^{\bar W} T^{[1]}_{\bar WY} \tilde{X}_c^Y \tilde{X}_k^Y \right. \nonumber \\
% - BC,KI CJ,AK
& \left. - \sum_{c\bar k} \sum_{T\bar U\bar WY} \tilde{\tilde{X}}_a^T \tilde{\tilde{X}}_c^T V_{T\bar U} X_{\bar j}^{\bar U} X_{\bar k}^{\bar U} \tilde{X}_{\bar b}^{\bar W} \tilde{X}_{\bar k}^{\bar W} T^{[1]}_{\bar WY} \tilde{X}_c^Y \tilde{X}_i^Y \right] \nonumber \\
&+ \sum_{\bar a\bar b\bar i\bar j} \sum_{\bar R\bar S} \left( \tilde{X}_a^R \tilde{X}_i^R T^{[1]}_{\bar R\bar S} \tilde{X}_b^S \tilde{X}_j^S - \tilde{X}_a^R \tilde{X}_b^R T^{[1]}_{\bar R\bar S} \tilde{X}_b^S \tilde{X}_i^S \right) \nonumber \\
% IK,JL AK,BL
&\times \left[ \frac{1}{2} \sum_{\bar k\bar l} \sum_{\bar T\bar U\bar W\bar Y} X_{\bar k}^{\bar T} X_{\bar i}^{\bar T} V_{\bar T\bar U} X_{\bar l}^{\bar U} X_{\bar j}^{\bar U} \tilde{X}_{\bar a}^{\bar W} \tilde{X}_{\bar k}^{\bar W} T^{[1]}_{\bar W\bar Y} \tilde{X}_{\bar b}^{\bar Y} \tilde{X}_{\bar l}^{\bar Y} \right. \nonumber \\
% AC,BD CI,DJ
& \left. + \frac{1}{2} \sum_{\bar c\bar d} \sum_{\bar T\bar U\bar W\bar Y} \tilde{\tilde{X}}_{\bar a}^{\bar T} \tilde{\tilde{X}}_{\bar c}^{\bar T} V_{\bar T\bar U} \tilde{\tilde{X}}_{\bar b}^{\bar U} \tilde{\tilde{X}}_{\bar d}^{\bar U} \tilde{X}_{\bar c}^{\bar W} \tilde{X}_{\bar i}^{\bar W} T^{[1]}_{\bar W\bar Y} \tilde{X}_{\bar d}^{\bar Y} \tilde{X}_{\bar j}^{\bar Y} \right. \nonumber \\
% AI,KC CK,BJ
& \left. + \sum_{\bar c\bar k} \sum_{\bar T\bar U\bar W\bar Y} \tilde{X}_{\bar c}^{\bar T} \tilde{X}_{\bar k}^{\bar T} V_{\bar T\bar U} \tilde{X}_{\bar b}^{\bar U} \tilde{X}_{\bar j}^{\bar U} \tilde{X}_{\bar a}^{\bar W} \tilde{X}_{\bar i}^{\bar W} T^{[1]}_{\bar W\bar Y} \tilde{X}_{\bar c}^{\bar Y} \tilde{X}_{\bar k}^{\bar Y} \right. \nonumber \\
% AI,kc ck,BJ
& \left. + \sum_{ck} \sum_{T\bar U\bar WY} \tilde{X}_{\bar c}^{\bar T} \tilde{X}_{\bar k}^{\bar T} V_{T\bar U} \tilde{X}_{\bar b}^{\bar U} \tilde{X}_{\bar j}^{\bar U} \tilde{X}_{\bar a}^{\bar W} \tilde{X}_{\bar i}^{\bar W} T^{[1]}_{\bar WY} \tilde{X}_c^Y \tilde{X}_k^Y \right. \nonumber \\
% - AI,KC CJ,BK
& \left. - \sum_{\bar c\bar k} \sum_{\bar T\bar U\bar W\bar Y} \tilde{X}_{\bar c}^{\bar T} \tilde{X}_{\bar k}^{\bar T} V_{\bar T\bar U} \tilde{X}_{\bar b}^{\bar U} \tilde{X}_{\bar j}^{\bar U} \tilde{X}_{\bar a}^{\bar W} \tilde{X}_{\bar k}^{\bar W} T^{[1]}_{\bar W\bar Y} \tilde{X}_{\bar c}^{\bar Y} \tilde{X}_{\bar i}^{\bar Y} \right. \nonumber \\
% - AC,KI CK,BJ
& \left. - \sum_{\bar c\bar k} \sum_{\bar T\bar U\bar W\bar Y} \tilde{\tilde{X}}_{\bar b}^{\bar T} \tilde{\tilde{X}}_{\bar c}^{\bar T} V_{\bar T\bar U} X_{\bar j}^{\bar U} X_{\bar k}^{\bar U} \tilde{X}_{\bar a}^{\bar W} \tilde{X}_{\bar i}^{\bar W} T^{[1]}_{\bar W\bar Y} \tilde{X}_{\bar c}^{\bar Y} \tilde{X}_{\bar k}^{\bar Y} \right. \nonumber \\
% - AC,KJ CI,BK
& \left. - \sum_{\bar c\bar k} \sum_{\bar T\bar U\bar W\bar Y} \tilde{\tilde{X}}_{\bar b}^{\bar T} \tilde{\tilde{X}}_{\bar c}^{\bar T} V_{\bar T\bar U} X_{\bar i}^{\bar U} X_{\bar k}^{\bar U} \tilde{X}_{\bar a}^{\bar W} \tilde{X}_{\bar k}^{\bar W} T^{[1]}_{\bar W\bar Y} \tilde{X}_{\bar c}^{\bar Y} \tilde{X}_{\bar j}^{\bar Y} \right]
\end{align}
These equations bear a striking similarity to those in the closed-shell case, where each open-shell term corresponds exactly to one of the closed-shell terms, except for numerical prefactor and the spin of each electron (loop). We leverage this similarity in our implementation by adding loops over $\alpha$ and $\beta$ spins to the closed-shell code, resulting in a highly efficient and maintainable implementation.

\textcolor{review}{In summary, the procedure for deriving open-shell THC methods is quite straightforward:
\begin{enumerate}
\item Enumerate all distinct Goldstone diagrams, for example by expanding the Brandow diagrams/ASGs via permutation of each vertex's indices.
\item Algebraically evaluate each diagram as usual,\cite{manybodymethods} except that the factor of 2 for closed loops is replace by an unrestricted sum over $\alpha$ and $\beta$ spin for each loop (with the same spin for all indices in the loop). For open diagrams (amplitude equations), the spin of external loops is fixed and determined by the spin of the corresponding amplitude to be determined.
\item Replace each two-electron integral and double excitation amplitude by its THC-factorized form. For open diagrams, additional collocation matrices are applied to the external lines in order to form the fitting matrix which will be used to determine the core matrix by least-squares fitting as in \eqref{eq:least-squares} and \eqref{eq:fitting}.
\end{enumerate}}

\section{Computational Details}

\begin{figure}[!h]
    \centering
    \includegraphics[width=0.6\textwidth]{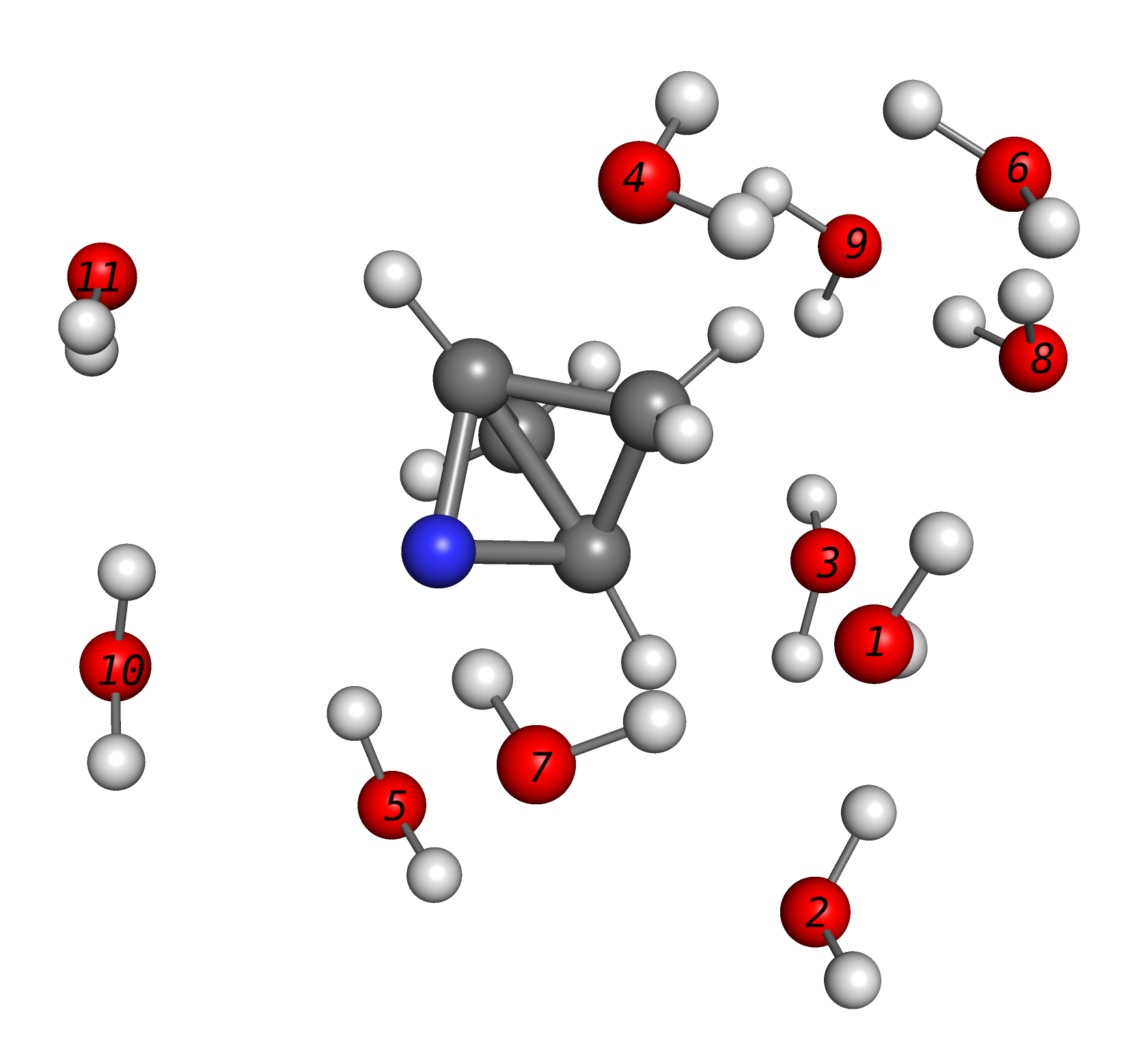}
    \caption{2H-2-azabicyclo[1.1.1]pentane radical with solvation shell. Solvent waters are numbered by center of mass distance from the solute.}
    \label{fig:pentane_solvation_shell}
\end{figure}

\begin{figure}[!h]
    \centering
    \includegraphics[width=0.8\textwidth]{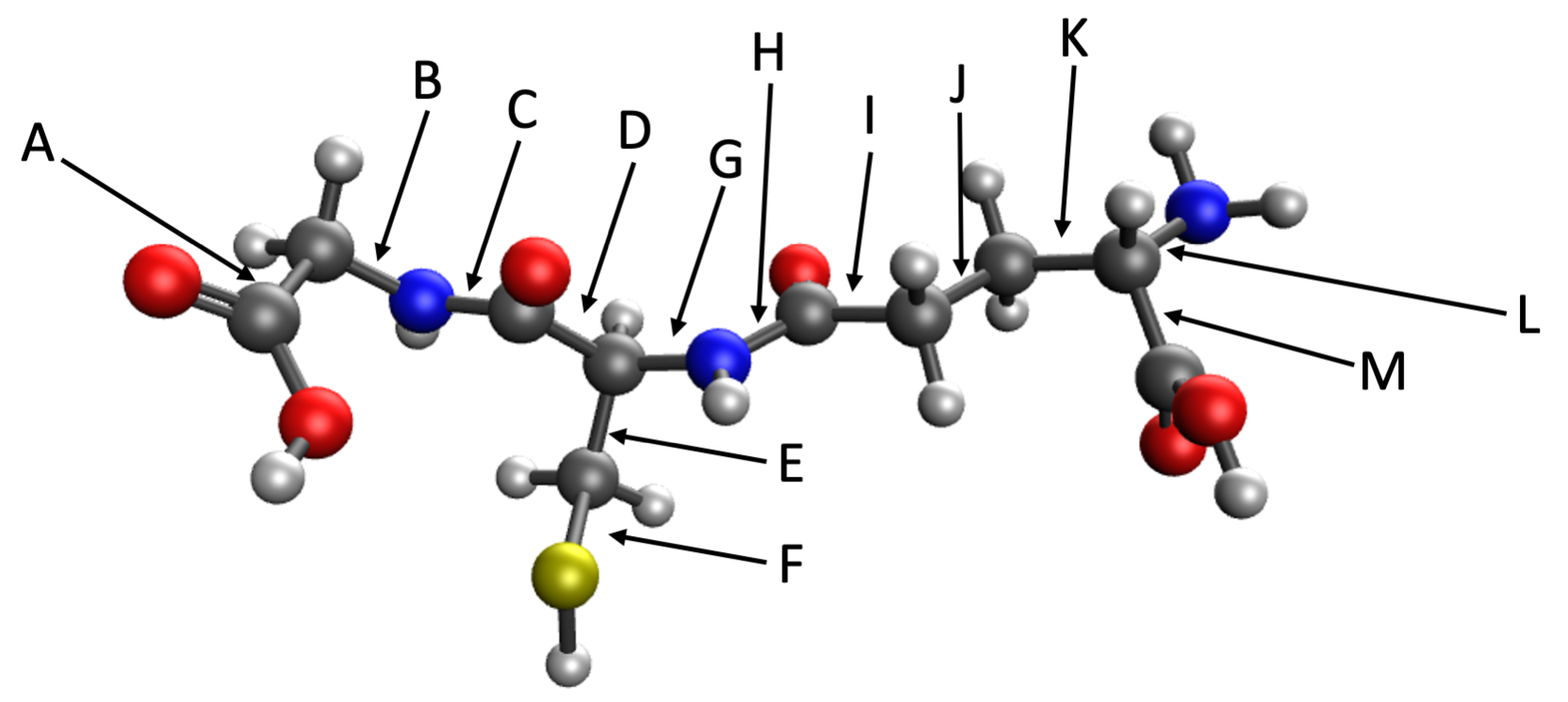}
    \caption{Ball-and-stick structure of glutathione with all backbone and \ce{C-S} bonds labeled.}
    \label{fig:glutathione_arrows}
\end{figure}

\begin{figure}[!h]
    \centering
    \includegraphics[width=0.6\textwidth]{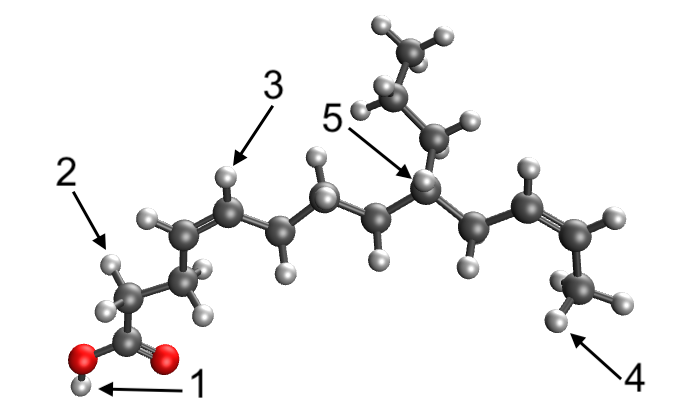}
    \caption{Ball-and-stick structure of 9-propyl-4,11-tridecadienoic acid with arrows pointing to specific hydrogens referenced in the text.}
    \label{fig:tridecadienoicacid_plucks}
\end{figure}

Open-shell LS-THC-MP2a, -MP2b, and -MP3b were implemented in a development version of CFOUR \cite{cfour}. We tested the accuracy of these methods on four types of test systems:
\begin{enumerate}
    \item \textbf{Linear alkyl radicals (\ce{H(CH2)_n^.}, $n$=1--20)}: Geometries were optimized at the B3LYP/def2-TZVP level \textcolor{review}{with D3 dispersion correction,\cite{becke1993,lyp1988,def2tzvpbasis, dispersion}} starting with synthetic structures with $R_{CC} = \SI{1.54}{\AA}$, $R_{CH} = \SI{1.1}{\AA}$, and tetrahedral angles. 
    \item \textbf{Micro-solvated 2H-2-azabicyclo[1.1.1]pentane radical} (Fig.~\ref{fig:pentane_solvation_shell}): The aqueous micro-solvation environment and solute radical geometry were determined from a short QM/MM simulation (see ESI for details). Up to the 11 closest water molecules (based on the distance of the water oxygen atom to the closest solute atom) were retained in the THC calculations. Solvation energies were computed without geometry relaxation or counterpoise corrections.
    \item \textbf{Glutathione} (Fig.~\ref{fig:glutathione_arrows}): A gas-phase structure for glutathione %(\ce{C10N3O6SH18})
    was optimized at the B3LYP/def2-TZVP level \textcolor{review}{with D3 dispersion correction}. Heterolytic and homolytic bond cleavage energies were calculated for each bond indicated in Fig.~\ref{fig:glutathione_arrows}. Isolated bond cleavage fragments were re-optimized at the same level of theory. The fragment charges after heterolytic cleavage were assigned based on the lowest-energy configuration.
    \item \textbf{9-propyl-4,11-tridecadienoic acid}  (Fig.~\ref{fig:tridecadienoicacid_plucks}): An initial structure was obtained at the same B3LYP/def2-TZVP level \textcolor{review}{with D3 dispersion correction}, as well as radical and ionic structures produced by removing, in turn, each hydrogen atom indicated in Fig.~\ref{fig:tridecadienoicacid_plucks} followed by reoptimization. We also generated 24 distinct conformations using the FRee Online druG conformation generation (FROG) tool.\cite{frog} We then removed the tertiary hydrogen (\#5) from each conformation and reoptimized using B3LYP/def2-TVZP.
\end{enumerate}

All B3LYP geometry optimizations were performed with Q-Chem.\cite{qchem} For all THC calculations, we used the cc-pVDZ basis set, \cite{dunningbasis} density fitting with the cc-pVDZ-RI auxiliary basis set, \cite{Feyereisen1993,dunlapdf,auxbasis} and SG0 \cite{sg0grid} as the parent grid. The parent grid was pruned as in Ref.~\citenum{matthewslsthc2020}, based on a numerical cutoff $\epsilon$ which was varied from $10^{-1}$ to $10^{-4}$ in each experiment. An unrestricted Hartree--Fock (UHF) reference was used in all cases.

In most cases, spin contamination of the UHF reference was negligible ($< 5\%$), although certain systems (e.g. some radical bond-breaking fragments of glutathione and the vinylic tridecanoic acid radical) exhibit moderate spin contamination, with values of $\langle\hat{S}^2\rangle$ as high as 0.99. We specifically avoid the use of a restricted open-shell Hartree--Fock (ROHF) reference, as these systems provide a test case for how the THC approximation is affected by spin contamination.

\section{Results}

\subsection{Size-extensivity of the error} \label{alkane_analysis}

\begin{figure}[!h]
    \centering
    \includegraphics[width=\textwidth]{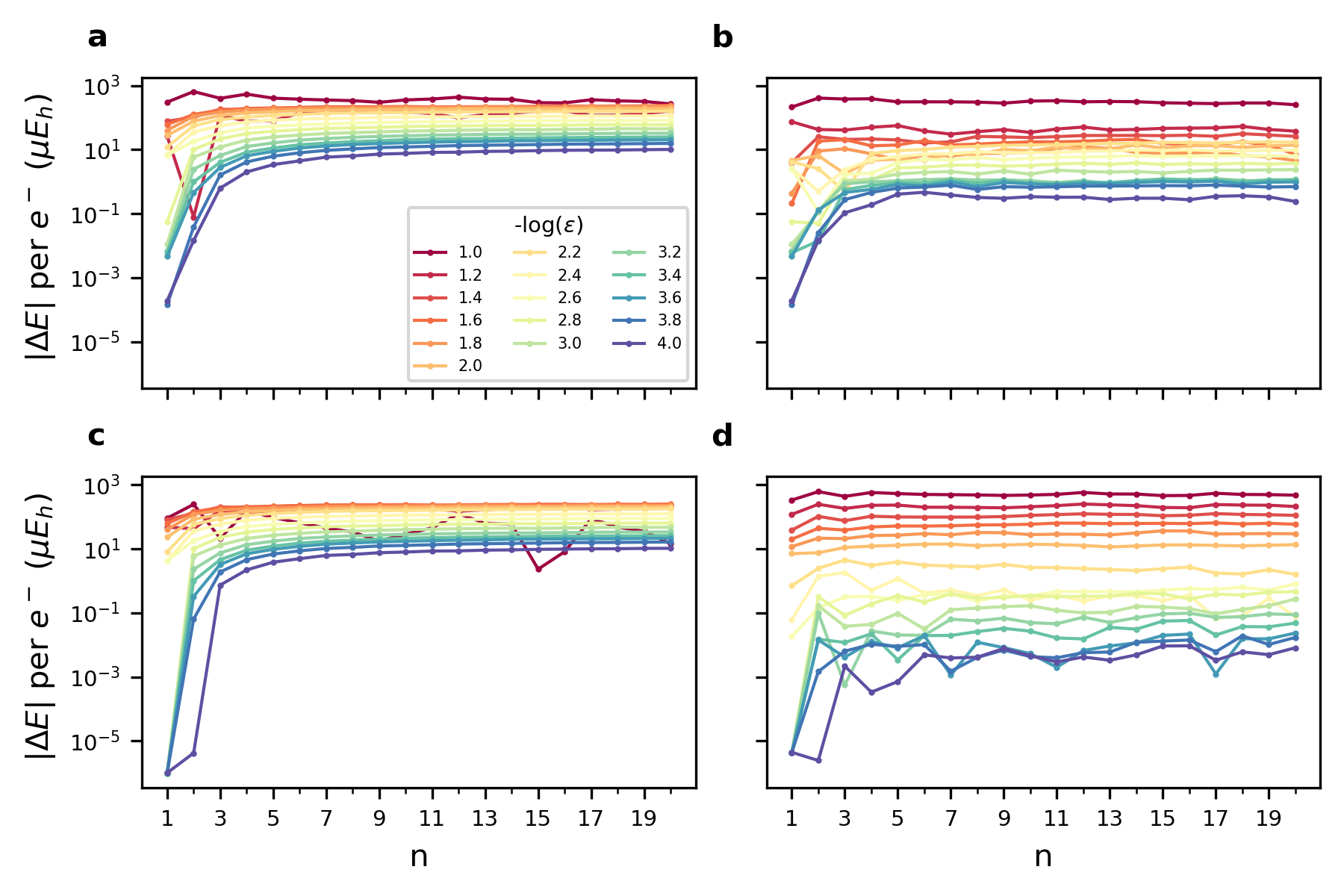}
    \caption{Absolute energy errors for open-shell linear alkyl radicals, \ce{H(CH2)_n^.}. The absolute value of the per-electron error between the THC and corresponding DF-MP$n$ calculation is plotted for \textbf{(a)} the total LS-THC-MP3b correlation energy, which is the sum of \textbf{(b)} the LS-THC-MP3b third-order correction and \textbf{(c)} the LS-THC-MP2b correlation energy, and finally \textbf{(d)} the LS-THC-MP2a correlation energy. \textcolor{review}{$\epsilon$ is the Cholesky decomposition cutoff parameter when pruning the grid, varying logarithmically from $10^{-1}$ to $10^{-4}$ in steps of 0.2 log units.}}
    \label{fig:alkane_ene_error}
\end{figure}

We first examine the error of the THC approximations compared to their density-fitted counterparts for linear alkyl radicals in order to correlate absolute energy errors with system size. Fig.~\ref{fig:alkane_ene_error} gives the THC error per correlated electron. THC calculations with a range of grid cutoff parameters, $\epsilon$, were performed in order to examine the dependence of the error with grid size (a smaller $\epsilon$ results in a larger grid and should yield a smaller error). The error per electron quickly reaches an approximately constant value, whether for the total MP3 correlation energy (Fig.~\ref{fig:alkane_ene_error}a), the MP2 correlation energy (Fig.~\ref{fig:alkane_ene_error}c,d), or the MP3 correction by itself (Fig.~\ref{fig:alkane_ene_error}b). The latter contribution is the most irregular, perhaps due to its smaller magnitude and sensitivity to the virtual-virtual grid represented by $\tilde{\tilde{\mathbf{X}}}$.

Notably, the error for LS-THC-MP2a is much smaller than for LS-THC-MP2b as previously observed in the closed-shell case. As the total LS-THC-MP3b correlation energy includes an LS-THC-MP2b contribution, this error is in fact dominated by the MP2 part, with the MP3 correction error being a minor part for most reasonable choices of $\epsilon$. In each case, the asymptotic error per electron decreases roughly linearly with respect to $\epsilon$ (note that the $\epsilon$ values are chosen on a log scale, and the $y$-axis is also logarithmic). Certain narrow ranges of $\epsilon$ significantly depart from this trend, e.g. $\epsilon \sim 10^{-1.2}$ in Fig.~\ref{fig:alkane_ene_error}a or $\epsilon \sim 10^{-3.0}$ in Fig.~\ref{fig:alkane_ene_error}d. These irregularities occur due to a sign change in the error.

\begin{comment}
Typical density-fitting errors for DF-MP3 seem to be approximately 15 $\mu E_h/e^-$ based on previous experiments[CITE]. This indicates that the LS-THC-MP2b errors are comparable for $\epsilon \sim XXXX$, or even earlier near $\epsilon \sim YYYY$ if the LS-THC-MP2b contribution is replaced by LS-THC-MP2a. These results confirm the size-extensivity of open-shell LS-THC-MP$n$ as was observed for the closed-shell variant. ERROR COMPARISON TO RHF?? (previous paper) We also observe, as for closed-shell THC, a "threshold" effect where convergence of the certain incremental error is also only reached for sufficiently long chain. This effect diminishes with looser cutoff values (smaller grids), suggesting a saturation of the orbital pair space for smaller systems.
\end{comment}

Typical density-fitting errors for DF-MP3 seem to be approximately 15 $\mu\mathrm{E_h}/e^-$ based on previous experiments. \cite{matthewslsthc2021}. This indicates that the LS-THC-MP2b errors are comparable for $\epsilon \sim 10^{-3.2}$, or even earlier near $\epsilon \sim 10^{-2.0}$ if the LS-THC-MP2b contribution is replaced by LS-THC-MP2a. These results confirm the size-extensivity of open-shell LS-THC-MP$n$ as was observed for the closed-shell variant. Typical per-electron errors are also similar for the closed- and open-shell THC methods; e.g. for \ce{C8H18}/\ce{C8H17^.} at $\epsilon = 10^{-2.4}$ we find errors of 1.0/0.35 $\mu\mathrm{E_h}/e^-$ (MP2a), 102/105 $\mu\mathrm{E_h}/e^-$ (MP2b), and 6.9/8.7 $\mu\mathrm{E_h}/e^-$ (MP3b correction), respectively. We also observe, as for closed-shell THC, a "threshold" effect where convergence of the incremental error to the asymptotic value is only reached for a sufficiently long chain. This effect diminishes with looser cutoff values (smaller grids), suggesting a saturation of the orbital pair space for smaller systems.

\subsection{Radical micro-solvation energies}

\begin{figure}[!h]
    \centering
    \includegraphics[width=\textwidth]{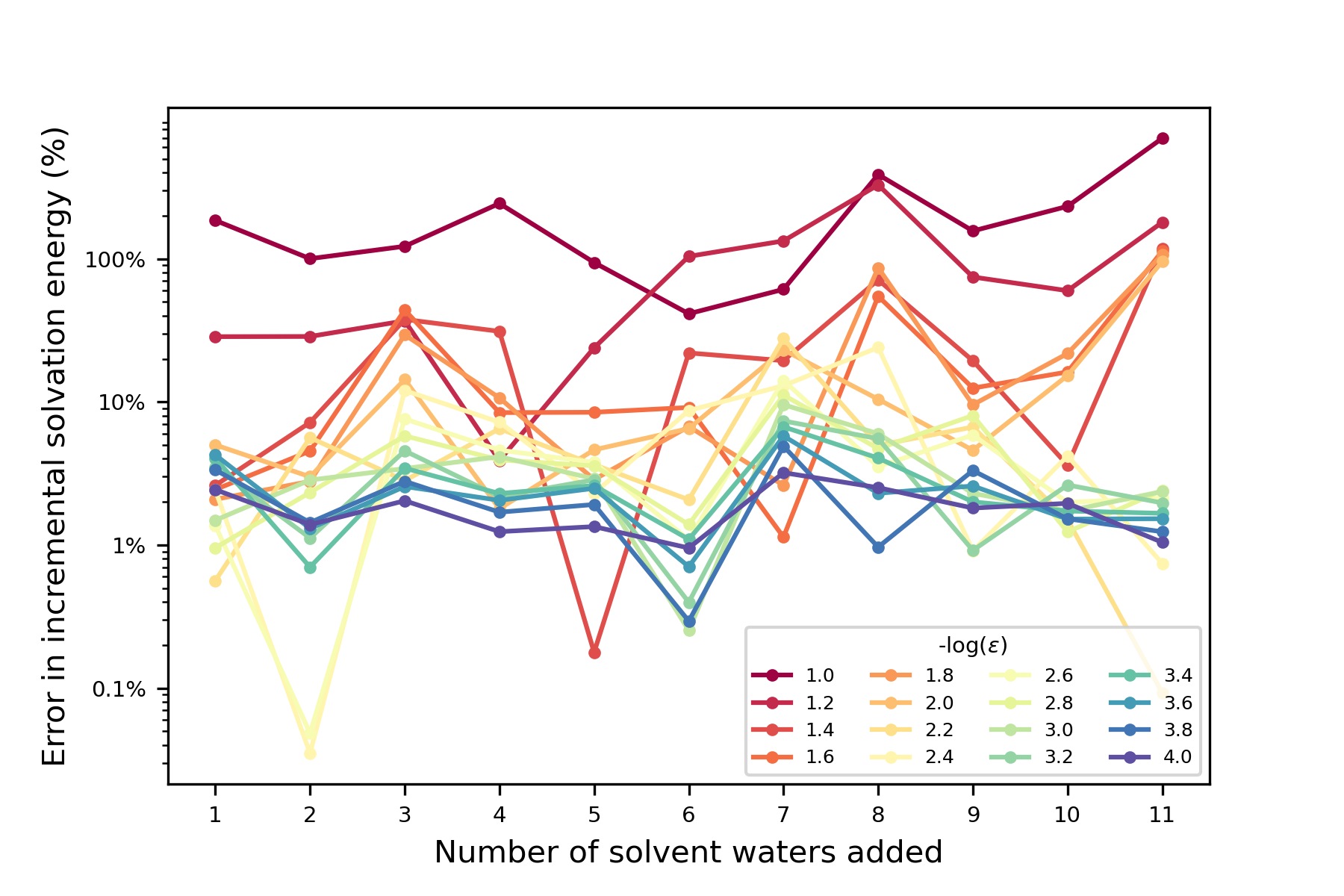}
    \caption{Errors in incremental solvation energy for 2H-2-azabicyclo[1.1.1]pentane in an explicit water solvation environment. From 1 to 11 waters are added in the order indicated in Fig.~\ref{fig:pentane_solvation_shell}. See text for details.}
    \label{fig:pentane_mp3}
\end{figure}

We next examine the error of the THC approximations for 2H-2-azabicyclo[1.1.1]pentane (ABP) in order to study how an increase in \textcolor{review}{solvation shell size} impacts solvation energy errors. This is an important test given that subsequent solvent waters will contribute very different physical interactions to the total solvation (interaction) energy. For example, the 5th and 7th waters added (see Fig.~\ref{fig:pentane_solvation_shell}) interact directly with the radical center, and the 7th water at least forming a hydrogen bond with the amino nitrogen. Other waters interact via weaker electrostatic interactions or van der Waals interactions and instead hydrogen bond with other solvent molecules (e.g. the 6th, 8th, and 9th waters), and some more distant waters do not seem to form any strong interactions, at least with other fragments included in the present calculations (e.g. the 10th and 11th waters). However, the errors due to THC, presented as percentages of the incremental solvation energy,
\begin{align}
   \%\Delta E_n &= \left| 1 - \frac{E_{THC-MP3}(\ce{ABP.(H2O)_n})-E_{THC-MP3}(\ce{ABP.(H2O)_{n-1}})}{E_{DF-MP3}(\ce{ABP.(H2O)_n})-E_{DF-MP3}(\ce{ABP.(H2O)_{n-1}})} \right|\times 100\%
\end{align}
do not show any clear trend with the strength or type of solvent interaction. Rather, the errors are relatively consistent at moderate cutoff values ($0.001<\epsilon<0.01$). At or below $\epsilon = 10^{-2.6}$, errors are consistently less than 5\%. The absence of a trend or significant outliers indicates a relative insensitivity of THC-MP3 to different types of interactions, and critically, no bias of the open-shell THC-MP3 method toward or against interactions involving unpaired electrons.

\subsection{Bond cleavages and radical stability}

\begin{figure}
    \centering
    \includegraphics[width=\textwidth]{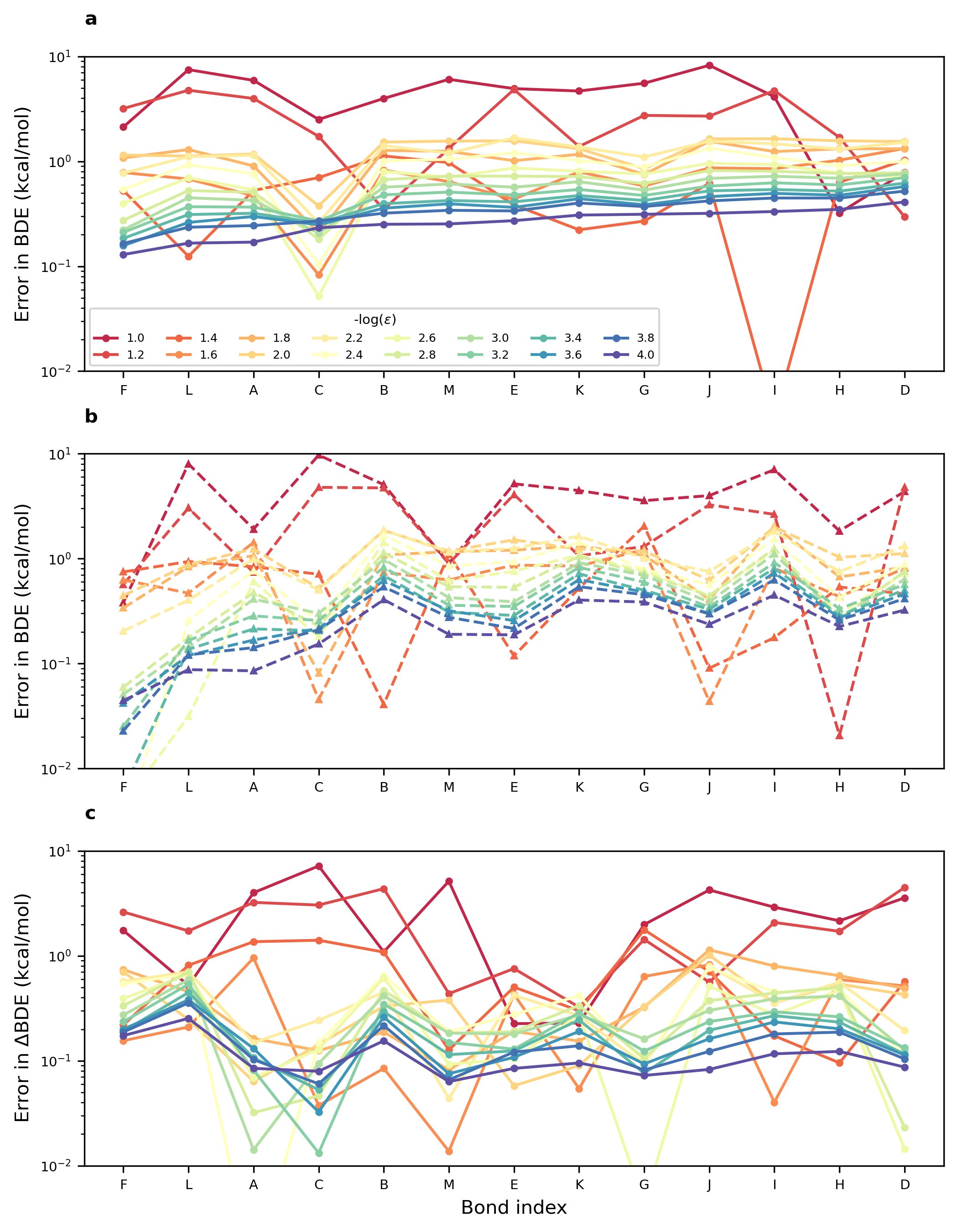}
    \caption{Bond dissociation energy errors of glutathione. \textcolor{review}{ \textbf{(a)} Absolute BDE error due to THC for homolytic bond cleavage (open-shell).  The bond index refers to labeled bonds in Fig. \ref{fig:glutathione_arrows}. \textbf{(b)} Absolute BDE error due to THC for heterolytic bond cleavage (closed-shell).} \textbf{(c)} Absolute errors in the difference between the homolytic and heterolytic BDEs due to THC.}
    \label{fig:Glutathione_mp3}
\end{figure}

We evaluate the robustness of the THC approximation with respect to various bond cleavage points and charge/spin separation by examining 13 different bond dissociation energies (BDEs) for the glutathione system (Fig.~\ref{fig:glutathione_arrows}) and 5 different hydrogen atom abstraction (HA) energies for the tridecadienoic acid system (Fig.~\ref{fig:tridecadienoicacid_plucks}).

We only consider bond breakages in glutathione between backbone \ce{C-C} or \ce{C-N} bonds, as well as the \ce{C-S} bond in the cysteine peptide. For each bond cleavage, we calculate both a homolytic ($\ce{AB} \rightarrow \ce{A^.} +\ce{B^.}$) and a heterolytic ($\ce{AB} \rightarrow \ce{A-} +\ce{B+}$ or $\ce{AB} \rightarrow \ce{A+} +\ce{B-}$, whichever results in more stable products) bond dissociation energy.

The homolytic and heterolytic bond dissociation energy errors with respect to standard density fitting calculations are shown in Fig.~\ref{fig:Glutathione_mp3}a. Overall, the THC approximation produces an accurate bond cleavage energy (around 1 kcal/mol) for most of the bond-breaking cases at reasonable selections of $\epsilon$ ($\epsilon < 0.01$). Additionally, the error differences between the homolytic (solid lines) and heterolytic (dashed lines) bond cleavage energy are of similar magnitude for each value of $\epsilon$, and at tighter thresholds follow the same trend towards lower \textcolor{review}{errors for bonds A, F, and L}. These bonds cleave the thiol or other terminal functional groups---because of the "threshold" effect noted above, these bond dissocations result in the lowest total error due to near-saturation of the THC grid for the smaller fragment. The closed-shell calculations result in smaller errors for these three cases, perhaps indicating a slight difference in how quickly the grid saturates in the open- and closed-shell cases. However, "moderate" cutoff values do not show such a trend and instead provide a rather consistent magnitude of error irrespective of the bond being broken.

The error in the difference between the homolytic and heterolytic BDEs ($\Delta$BDE) were also computed and are depicted in Fig.~\ref{fig:Glutathione_mp3}b. Across the various bonds, errors for this relative measure are somewhat lower than for the BDEs themselves. This indicates a reliable error cancellation between the open- and closed-shell LS-THC calculations on similar systems (note that the geometries of the products were optimized separately for hetero- and homolytic cleaveages).

\begin{figure}[!h]
    \centering
    \includegraphics[width=\textwidth]{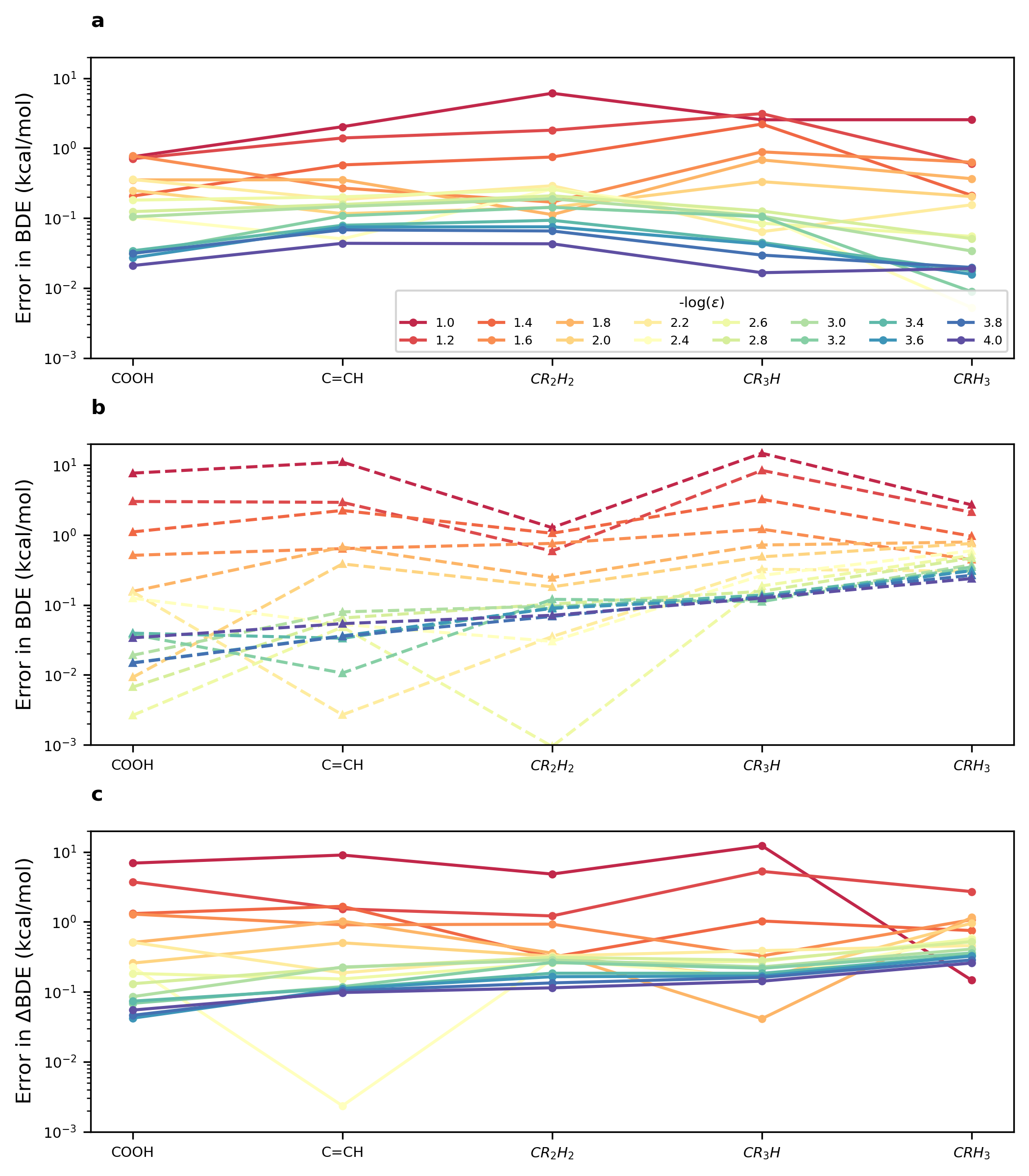}
    \caption{Bond dissociation energy errors of 9-propyl-4,11-tridecadienoic acid.  
    \textcolor{review}{\textbf{(a)} Absolute BDE error due to THC for homolytic bond cleavage (open-shell). \textbf{(b)} Absolute BDE error due to THC for heterolytic bond cleavage (closed-shell). \textbf{(c)} Absolute errors in the difference between the homolytic and heterolytic BDEs due to THC.}}
    \label{fig:Tridecadienoicacid_mp3_BDE}
\end{figure}

In Fig.~\ref{fig:Tridecadienoicacid_mp3_BDE}, bond dissociation energy errors are shown for $\mathrm{H}^{\{+,\bullet,-\}}$ abstraction from 9-propyl-4,11-tridecadienoic acid. For the closed-shell products (dashed lines; these are all cationic with the exception of the carboxylate) there does seem to be a trend which tends to result in lower errors for the more stable cationic products (substituted alkyl radicals and especially doubly-bonded sp$^2$ cationic centers). This may result from enhanced error cancellation between more similar geometries where re-hybridization is incomplete. The errors for radical open-shell products (solid lines) are more consistent, perhaps again due to reduced rehybridization even in the primary and secondary carbon radicals. Fig.~\ref{fig:Tridecadienoicacid_mp3_BDE}b gives the error in the relative energy between the charged and neutral abstraction products. As for glutathione, there is some cancellation of errors for, in particular for the carboxyl and vinyl abstractions where the geometric changes are more similar.

\begin{comment}
The second illustration of the bond cleavage energy errors is the 9-propyl-4,11-tridecadienoic acid system. It is of interest as it contains a long hydrocarbon chain, a beta carbon, primary and tertiary carbons, carboxylic acid and propyl functional groups, and double bonds between carbons, which all lead to interesting radicals.
The bond cleavage energy errors of the 9-propyl-4,11-tridecadienoic acid system are shown in Figure \ref{fig:Tridecadienoicacid_mp3}. The errors are negligible below 1 kcal/mol for all of the bond-breaking cases at a rather large tolerance ($\epsilon = 10^{-2}$). The correlation between energy error and radical stability is more obvious here. The cleavage that results in \ce{R-COO^.} is the most stable because of the resonance effect. The tertiary carbon radical is more stable than the primary carbon radical, and the two delocalization centers next to the secondary carbon radical (label 2), allow for this secondary carbon radical to be more stable than the tertiary carbon. The cleavage that was on one side of the carbon-carbon double bond (label 3) is the second most stable, most likely due to the double bond being made up to one sigma and one pi bond and the carbons on either side being of $sp^2$ hybridization. There is an outlier in the energy error where it is much smaller when the tolerance is $\epsilon = 10^{-2.4}$, most likely due to the double bond cleavage being most stable at that tolerance. 
\end{comment}

\subsection{Conformational energy ordering}

\begin{figure}[!h]
    \centering
    \includegraphics[width=\textwidth]{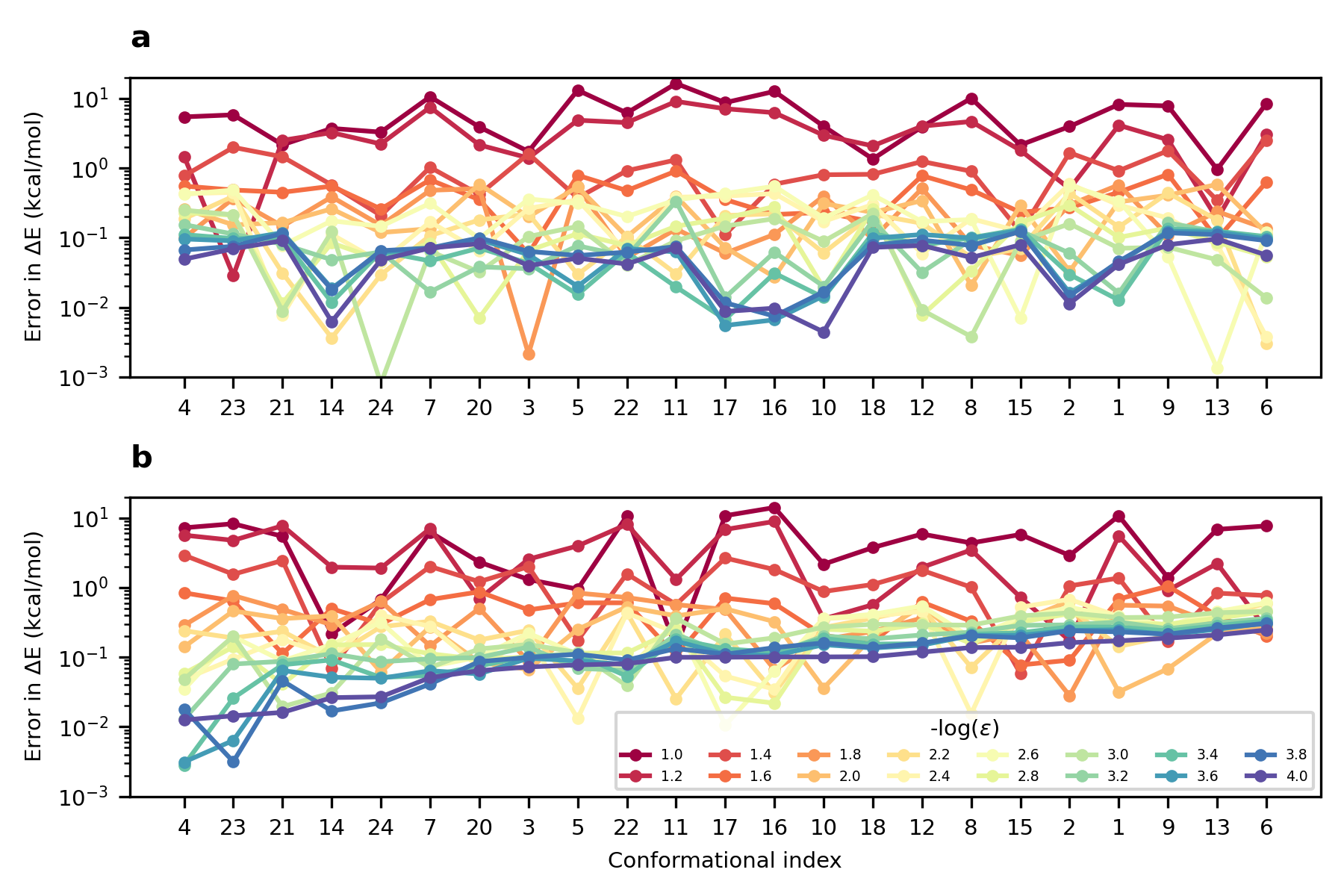}
    \caption{Errors in the relative energies of 24 conformations 9-propyl-4,11-tridecadienoic acid due to THC. Conformational energy differences are calculated with respect to the lowest energy conformation (index 19), and the remaining conformations are ordered by increasing energy (using the DF-MP3 energies). \textbf{(a)} Errors for open-shell radical conformations. \textbf{(b)} Errors for closed-shell cationic conformations.}
    \label{fig:conformation}
\end{figure}

Again focusing on 9-propyl-4,11-tridecadienoic acid, we examine the relative energies of 24 distinct conformations of the 9H hydrogen abstraction (open-shell radical) and hydride abstraction (cationic) products. In Fig.~\ref{fig:conformation}, the errors in relative conformational energies due to THC are presented for both types of products. While it is clear that there is not a strong correlation of the errors between the neutral radical and charged closed-shell products (c.f. the lack of any distinct trend in the open-shell errors with respect to the closed-shell errors ordered from smallest to largest at $\epsilon = 10^{-4}$). However, both the closed- and open-shell errors seem to span almost identical ranges for each value of $\epsilon$. The smallest and largest errors, even at a tight tolerance of $\epsilon = 10^{-4}$, seem to span at least an order of magnitude, although there does not seem to be a trend of the size of the error with any chemically-relevant features of the individual conformations. The fact that the smallest errors are significantly lower than observed in, for example, Fig.~\ref{fig:Glutathione_mp3} may then simply indicate fortuitious error cancellation which is not reproducible between the radical and cationic structures. Since the final structures in these cases were optimized independently (despite starting with the same algorithmically-generated guess structure), this is perhaps not surprising.

\begin{comment}
 with different conformations was evaluated by examining the cancellation of errors, where different angles and dihedral angles of the molecule were sampled. First, the absolute value of conformational energy differences with respect to the lowest energy conformation (index 19) was calculated, and the absolute value of the conformational energy differences errors with respect to DF was calculated and plotted (Fig.\ref{fig:conformation}). For both the UHF system (radicals) and the RHF system, the conformation energy difference errors with respect to the DF calculation were within 1 kcal/mol  starting at tolerance $ \epsilon = 10^{-1.6}$. Evidently, the errors obtained with tolerance $10^{-3.0}$ are indistinguishable (less than 0.1 kcal/mol) from the DF calculation. A systematic convergence pattern towards the exact value is also notable.
\end{comment}

\subsection{Computational scaling}

\begin{figure}[h!]
    \centering
    \includegraphics[width=\textwidth]{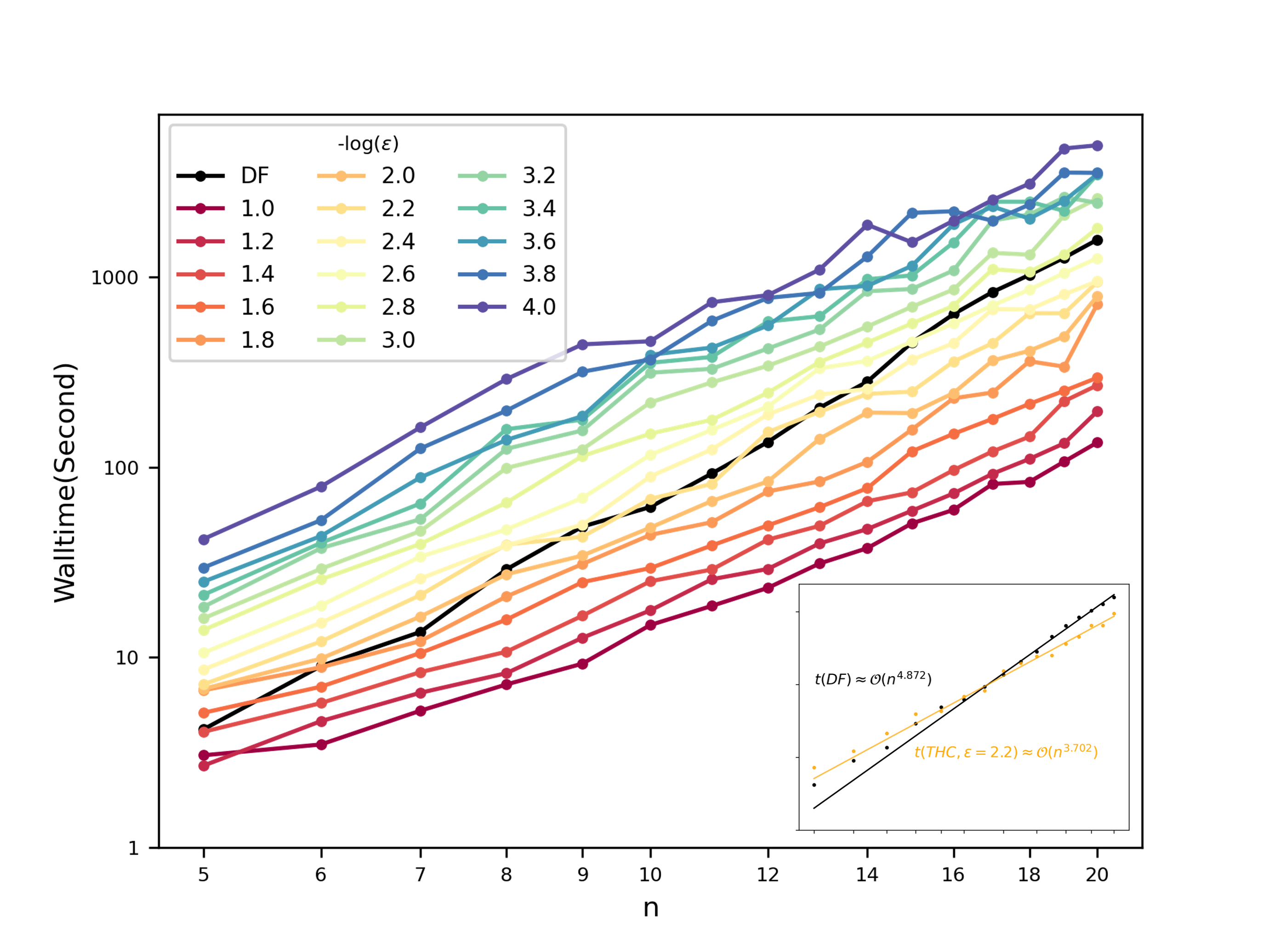}
    \caption{\textcolor{review}{Timings for DF- and LS-THC-MP3 methods for linear alkyl radicals, (\ce{H(CH2)_n^.}, $n$=5--20) with the cc-pVDZ basis set. Both axes are on a logarithmic scale to highlight the polynomial scaling. All calculations were performed using a single node with $2\times$ Intel Xeon E5-2695v4 CPUs and 256 GiB of memory; OpenMP was used to parallelize the calculation over all 36 cores. The inset shows the measured scaling of DF-MP3 and LS-THC-MP3 with $\epsilon=10^{-2.2}$ obtained via linear regression of the timings for $n=10$---20.}}
    \label{fig:alkane_timings}
\end{figure}

\textcolor{review}{In the above sections, we have shown that LS-THC-MP3 can achieve below 1 kcal/mol error with moderate cutoff values ($\epsilon \le 10^{-2.2}$). It is equally important to demonstrate the scaling reduction that can be achieved by the LS-THC method. To that end, we report the timings of both DF-MP3 and LS-THC-MP3 methods for linear alkyl radicals with varying cutoff parameters ($\epsilon$), analyzed in Section~\ref{alkane_analysis}\emph{Size-extensivity of the error}. The timings include the calculation of the MP2 and MP3 correlation energy only. As can be seen in Fig.~\ref{fig:alkane_timings}, is clear that LS-THC-MP3 achieves reduced scaling compared to DF-MP3 (see inset for measure scaling), and reaches a crossover at around 70 correlated electrons with a cutoff of $\epsilon = 10^{-2.2}$. Note that the DF-MP3 scaling in this regime is actually dominated by the formation of the $(ab|cd)$ integrals which scales as $\mathcal{O}(n^5)$, and other lower-scaling operations.}

\section{Conclusions}

The LS-THC method has proven to be an efficient approximation of both the ERI and doubles amplitudes tensors with high accuracy and low scaling for closed-shell systems.

Here, LS-THC was implemented for MP2 and MP3 calculations on open-shell systems and was evaluated with various test systems: linear alkyl radicals, a micro-solvated amine radical, heterolytic and homolytic bond cleavages in glutathione, and proton/hydrogen/hydride abstractions as well as conformational differences of 9-propyl-4,11-tridecadienoic acid. A number of trends were observed:
\begin{enumerate}
\item Similar to the RHF implementation, the LS-THC-MP$n$ errors scale linearly with system size, after a critical "threshold" molecular size is reached (this effect is diminished for looser cutoff values). Errors for LS-THC-MP2a are essentially negligible, while errors in LS-THC-MP3b are dominated by the MP2b contribution. Remaining errors in only the LS-THC-MP3b contribution are reasonably small with a double-zeta basis set.
\item Reliable error cancellation is evident in almost all calculations of energy differences. The greatest degree of error cancellation occurs when the systems are most chemically similar, resulting fragments are unequally distributed in size, or potentially when more stable products are produced. Errors in relative energies are typically below 1 kcal/mol for moderate cutoff values in the range $\epsilon < 10^{-2.2}$.
\item The errors produced for open-shell systems are highly similar to the errors produced for chemically similar closed-shell systems. In some cases, error cancellation can also be exploited between open- and closed-shell processes, such as in the relative BDEs of heterolytic and homolytic bond cleavages.
\item The error of the open-shell LS-THC-MP$n$ methods is highly insensitive to the specific nature of the chemical structure, type of interactions, and even moderately severe spin contamination of the reference wavefunction.
\end{enumerate}

In summary, \textcolor{review}{\emph{open-shell THC seems to be equally as applicable as closed-shell THC and multi-reference THC}\cite{Song2018,Song2020}.}
The diagrammatic method of derivation presented also enables the implementation of open-shell THC methods with little additional effort compared to the closed-shell version, and a highly similar code structure which should enable maintainable, efficient codes.

\section*{Acknowledgements}

This work was supported by the US Department of Energy under grant DOE-SC0022893, and in part by the US National Science Foundation under grant OAC-2003931 and CHE-2143725. MS is supported by an SMU Center for Research Computing Graduate Fellowship. All calculations were performed on the ManeFrame II computing system at SMU.

\section*{Disclosure Statement}

No potential conflict of interest was reported by the authors.

\section*{Supporting Information}

The following electronic supplementary information files are available from the publisher's website:
\begin{itemize}
\item The factorized LS-THC MP3b equations and QM/MM simulation details (.pdf).
\item All calculated DF- and LS-THC-MP$n$ correlation energies (.xlsx).
\item Molecular geometries used for all calculations (.xlsx).
\end{itemize}

\bibliography{references.bib}

\end{document}